\documentclass[a4paper,fleqn]{cas-dc}

\usepackage[T1]{fontenc}
\usepackage{graphicx}
\usepackage{arydshln}
\usepackage{graphbox}
\usepackage{algorithm}
\usepackage[noend]{algpseudocode}
\usepackage{amsfonts}
\usepackage{amsmath}
\usepackage{array}
\usepackage{mathtools}
\usepackage{xcolor}
\usepackage[numbers]{natbib}
\usepackage{enumitem}
\setlist[itemize]{leftmargin=*}

\usepackage{mathtools}
\usepackage{stmaryrd}
\usepackage{amssymb}
\usepackage{moreverb}
\usepackage{amssymb}
\usepackage{lineno}
\usepackage{graphicx}
\usepackage{amsmath}
\usepackage{moreverb}
\usepackage{amssymb}
\usepackage{lineno}
\usepackage{graphicx}
\graphicspath{ {images/} }
\usepackage{multirow}
\usepackage[utf8]{inputenc}
\usepackage{mathtools}
\let\oldnl\nl
\newcommand{\nonl}{\renewcommand{\nl}{\let\nl\oldnl}}
\usepackage{stackengine}
\usepackage{float}
\usepackage{stmaryrd}
\usepackage{multirow}
\usepackage{epstopdf}
\usepackage{makecell}
\usepackage{multirow}
\usepackage{graphicx}
\usepackage[colorlinks]{hyperref}
\usepackage{lipsum}
\usepackage{setspace}
\usepackage[english]{babel}
\usepackage{microtype}
\begin{document}
	\let\WriteBookmarks\relax
	\def\floatpagepagefraction{1}
	\def\textpagefraction{.001}
	\shorttitle{MicroFog Framework}
	\shortauthors{S. Pallewatta et~al.}
	
	\title [mode = title]{MicroFog: A Framework for Scalable Placement of Microservices-based IoT Applications in Federated Fog Environments}

	\author{Samodha Pallewatta}
        \ead{ppallewatta@student.unimelb.edu.au }

	\author{Vassilis Kostakos}

	\author{Rajkumar Buyya}

	\address{The Cloud Computing and Distributed Systems (CLOUDS) Laboratory, School of Computing and Information Systems, The University of Melbourne, Australia}

	
	
	\begin{abstract}
MicroService Architecture (MSA) is gaining rapid popularity for developing large-scale IoT applications for deployment within distributed and resource-constrained Fog computing environments. As a cloud-native application architecture, the true power of microservices comes from their loosely coupled, independently deployable and scalable nature, enabling distributed placement and dynamic composition across federated Fog and Cloud clusters. Thus, it is necessary to develop novel placement algorithms that utilise these microservice characteristics to improve the performance of the applications. However, existing Fog computing frameworks lack support for integrating such placement policies due to their shortcomings in multiple areas, including MSA application placement and deployment across multi-fog multi-cloud environments, dynamic microservice composition across multiple distributed clusters, scalability of the framework to operate within federated environments, support for deploying heterogeneous microservice applications, etc. To this end, we design and implement MicroFog, a Fog computing framework compatible with cloud-native technologies such as Docker, Kubernetes and Istio. MicroFog provides an extensible and configurable control engine that executes placement algorithms and deploys applications across federated Fog environments. Furthermore, MicroFog provides a sufficient abstraction over container orchestration and dynamic microservice composition, thus enabling users to easily incorporate new placement policies and evaluate their performance. The capabilities of the MicroFog framework, such as the scalability and flexibility of the design and deployment architecture of MicroFog and its ability to ensure the deployment and composition of microservices across distributed fog-cloud environments, are validated using multiple use cases. Experiments also demonstrate MicroFog's ability to integrate and evaluate novel placement policies and load-balancing techniques. To this end, we integrate multiple microservice placement policies to demonstrate MicroFog's ability to support horizontally scaled placement, service discovery and load balancing of microservices across federated environments, thus reducing the application service response time up to 54\%.

	\end{abstract}
	
	\begin{keywords}
		Fog Computing \sep Microservices\sep Application Placement\sep Internet of Things \sep Microservice Composition
	\end{keywords}

	\maketitle
	\section{Introduction}
        The Internet of Things (IoT) is growing rapidly, and the ever-increasing number and variety of connected devices generate massive amounts of data related to a wide range of smart application domains, such as smart cities, smart healthcare, Industrial IoT, and smart transportation, to mention a few. Hence, IoT application development is adapting Microservices Architecture (MSA) to support the rapid evolution of IoT application development towards creating an IoT ecosystem. Being a cloud-native application architecture, MSA builds applications as collections of modules known as microservices that are independently deployable and scalable \cite{websiteFowler}. Microservices are containerised using technologies such as Docker and dynamically composed using container orchestration platforms like Kubernetes and service mesh technologies such as Istio, thus ensuring seamless connectivity among microservices deployed across distributed computing resources. 

        Meanwhile, Fog computing is emerging as a powerful distributed computing paradigm for hosting latency-critical and bandwidth-hungry IoT applications. Fog computing extends cloud-like services towards the edge of the network by using the computing, networking and storage resources residing within the path connecting IoT devices to the centralised Cloud data centres \cite{mahmud2018fog}. With the increasing use of IoT applications, sending large amounts of data towards the centralised Cloud incurs high latency and bandwidth congestion. Moreover, distributed Fog resources provide the location awareness, data security, mobility awareness, and scalability required by the IoT applications, with geo-distributed users seeking ubiquitous access to the application services \cite{goudarzi2022scheduling}.  
        
        While the use of distributed Fog computing resources is a solution for this, the resource-constrained nature of the Fog resources is the main drawback which can be overcome through the federation of geo-distributed Fog clusters and Cloud data centres. This includes cooperative use of distributed Fog computing cluster/ data centres and Cloud data centres for the placement of applications to satisfy their demands and meet QoS requirements \cite{do2019systematic}. Such an approach focuses on extending the hybrid Cloud to include Fog computing resources provided by multiple Fog Infrastructure Providers (FIP) and maintain seamless connectivity across different environments to achieve the best possible performance \cite{farzin2022flex}. Furthermore, cloud-native characteristics of microservices make them perfect for such placement of large-scale IoT applications, which has given rise to novel paradigms like Osmotic Computing that proposes the convergence of IoT, MSA and Fog computing where microservices are dynamically moved and composed across hybrid fog-cloud environments \cite{neha2022systematic}.

        To harvest the full potential of MSA in Fog computing environments, the development of efficient placement algorithms is of vital importance. Thus, research on designing, developing and evaluating algorithms for the placement of microservice-based IoT applications is attracting a lot of attention. Existing literature contains works focusing on horizontally scaled placement of microservices to meet QoS parameters such as throughput, reliability and latency \cite{faticanti2019cutting, guerrero2019evaluation, deng2020optimal,pallewatta2022qos}, location-aware placements \cite{guo2022joint}, etc. that place interconnected microservices across distributed resources. However, these algorithms require extensive and accurate evaluations and validations before applying them at the enterprise level \cite{mahmud2022ifogsim2}. 

        Evaluation of the placement policies can be conducted using numerical evaluations \cite{guo2022joint,guerrero2019evaluation}, simulators \cite{fang2020iot, pallewatta2022qos, paul2020crew} and real-world deployments through small-scale testbeds \cite{faticanti2019cutting, fu2021qos}. Cloud computing-related policy evaluation can be conducted using Cloud infrastructure provided by commercial service providers like Amazon AWS, Google Cloud, etc., through their Infrastructure as a Service (Iaas) offerings through a rental model. Due to the lack of such platforms for Fog computing, Fog application placement policies are primarily evaluated using numerical evaluations and simulators. Although several real-world frameworks are available to manage Fog resources \cite{deng2021fogbus2, santoro2017foggy}, they have limitations related to Microservices-based IoT application placement. They lack support for the dynamic composition of microservices across federated Fog and Cloud data centres, easy integration of distributed placement policies, compatibility with open-source cloud-native technologies, support for heterogeneous microservices-based applications, ease of setup and prototyping support, etc. To overcome these limitations, we propose MicroFog: an easily configurable software framework for microservice-based application placement within federated fog-cloud environments. MicroFog can be used by IoT application developers, Fog infrastructure providers, and researchers in Fog computing to create, integrate and evaluate novel placement policies to deploy and manage microservices-based IoT applications. MicroFog enables the users to create placement approaches that harvest the potential of MSA, thus improving the QoS of applications.

        MicroFog provides a configurable control engine that executes placement policies in a distributed or centralised manner and deploys containerised microservices within Kubernetes and Istio-managed Fog and Cloud resource clusters. MicroFog abstracts Kubernetes and Istio resource deployment (i.e., pods, services, virtual services, gateways, etc.) while providing support for integrating novel placement algorithms and load-balancing policies. Moreover, MicroFog ensures the dynamic composition of microservices distributed across geo-distributed multi-fog multi-cloud environments by enabling service discovery and load balancing.

        The major contributions of our work are as follows:

        \begin{itemize}[leftmargin=*]
            \item A scalable and extensible framework is proposed for deploying and managing microservices-based IoT applications within the federated Fog and Cloud environments. The framework consists of multiple components, including a Control Engine (MicroFog-CE) for placement algorithms execution and application deployment, data stores to store required metadata, a monitoring component and a logging component.
            \item MicroFog-CE is designed and developed as an easy-to-configure microservice supporting different operation modes (centralised vs distributed), application placement modes (periodic vs event-driven), integration of novel placement policies, load balancing policies, etc.   
            \item Deployment architectures are proposed for the major components of the MicroFog framework to ensure their scalable and fault-tolerant deployment across federate Fog and Cloud environments.
            \item A proof-of-concept prototype of the framework is created, and the main features of the framework are demonstrated and evaluated using multiple use cases and benchmark policies integrated with the control engine. 
        \end{itemize}

The rest of the paper is organised as follows. In Section \ref{sec:background}, we provide a comprehensive background on microservices-based application placement, derive requirements of the framework based on that and analyse related research. Section \ref{sec:microfog} introduces the MicroFog framework, and Section \ref{sec:deployment} details the deployment architectures for the main components of the framework. APIs to access MicroFog-CE are presented in Section \ref{sec:apis}. Features of the framework are evaluated in section \ref{sec:evaluation}. Finally, Section \ref{sec:conclusion} concludes the paper. 

\section{Background and Related works} \label{sec:background}
In this section, we present a comprehensive background on the Fog computing paradigm, microservices-based applications, their deployment-related aspects and the Fog application placement problem to derive requirements of the frameworks for scalable Placement of Microservices-based IoT Applications within Federated Fog Environments. Moreover, we provide a qualitative comparison of existing frameworks to highlight the capabilities of our proposed framework.

\subsection{Fog Computing}\label{sec:fogmodel}

\begin{figure}[!ht]
    \includegraphics[width=\linewidth]{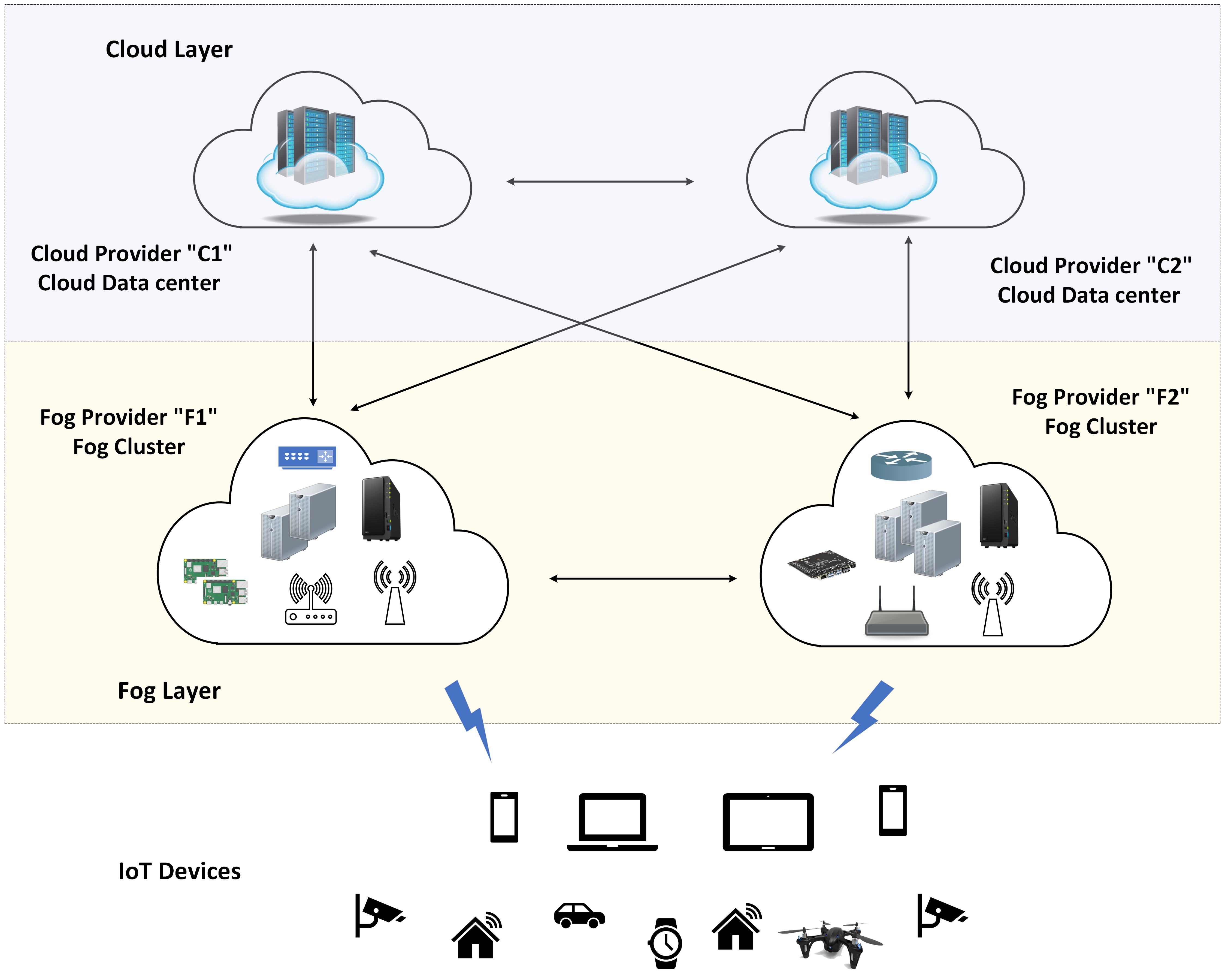}
    \caption{Federated multi-fog and multi-cloud Architecture}
    \label{fig:fogArchitecture}
    \vspace*{-2.0mm}
\end{figure}

Fog computing introduces an intermediate layer between IoT devices and the Cloud, consisting of distributed, heterogeneous and resource-constrained resources compared to Cloud data centres \cite{mahmud2018fog}. With the rapid growth in IoT applications, Fog computing is evolving towards a federated multi-fog multi-cloud architecture \cite{farzin2022flex} where multiple Fog Providers provide infrastructure, including computing, storage and networking resources within the Fog layer. This helps to overcome the resource-constrained natures of the Fog devices, enables ubiquitous access, and supports location-aware placement of applications. In this work, we consider the existence of multiple such Fog clusters provided by various service providers where they maintain connectivity with neighbouring clusters and the Cloud.

\subsection{Microservices-based  Applications}

MicroService Architecture (MSA) decomposes an application into a set of independently deployable modules known as microservices designed around business logic to have well-defined business boundaries \cite{joseph2019straddling}. Microservices communicate with each other using lightweight APIs to create composite services that the end users access. 

The loosely coupled nature of these microservices enables them to be deployed and scaled independently within distributed environments. Thus, dynamic service discovery and load-balancing mechanisms ensure seamless connectivity among microservices. To achieve such cloud-native behaviour, microservices are packaged as containers (i.e., Docker) that can be scaled (up and down) rapidly to meet the request demand. With such technologies, MSA can deploy microservices across distributed multi-fog multi-cloud environments while maintaining seamless connectivity and dynamic load balancing among horizontally scaled instances.

\subsubsection{Modelling Microservice Application}\label{sec:appModel}
As microservices-based applications have interactions among microservices, they can be modelled using Directed Acyclic Graphs (DAGs) \cite{pallewatta2022qos} where the vertices of the DAG represent microservices ( $m \in M_a$ where $M_a$ is the set of microservices of application $a$). Directed edges in DAG represent microservice invocations such that the direction is from the client microservice (consumer) to the invoked microservice (consumed). Microservices are independently packaged and have heterogeneous resource requirements that can be defined in terms of required RAM, CPU, storage, etc., needed to satisfy a specific request rate/throughput. Due to the fine-grained nature of the microservices, they communicate to create composite services where each application provides multiple services ($S_a$: the set of services of application $a$) with heterogeneous QoS requirements that can be defined at the service level. As microservices can have complex interaction patterns to create composite services (i.e., chained, aggregator, hybrid), the dataflows among microservices can be uni-directional or bi-directional ($df^a$: set of dataflows among $m \in M_a$). Thus, each application can be denoted as a tuple of $<M_a, df^a, S_a>$ where each service $s \in S_a$ is depicted by a tuple containing its microservices, data paths within them and QoS requirements of the service; $<M_a^s, P_a^s, Req_a$>. Data paths are collections of dataflows within a composite service that can be used to calculate the makespan of the service. It depends on the interaction pattern of the microservices within the composite service (i.e., the chained pattern has a single data path, whereas the aggregator invokes multiple datapaths). 
    
\subsection{ Application Deployment Related Aspects}
Microservices-based application deployment and management are aided by three cloud-native technologies: containerisation platforms (i.e., Docker), container orchestration systems (i.e., Kubernetes, Docker Swarm) and service mesh platforms (i.e., Istio, Consul). The MicroFog framework proposed in this work uses Docker, Kubernetes and Istio for the deployment and management of the microservices. Hence, we describe each technology and its aspects related to the federated fog-cloud deployment of applications as follows: 

\subsection{Containerisation using Docker} 
Microservices are packaged as Docker containers to make them independent of the host environments. Moreover, compared to earlier used virtual machines, containers are light-weigh with less startup time. Thus, containerisation of the microservices suits distributed deployment and scaling across heterogeneous and resource-constrained Fog nodes. Docker container images are stored and distributed using a container registry. Docker provides a fully managed container repository known as DockerHub. However, this is a centralised repository with limitations in privacy and security. Pulling images from a centralised repository can incur extra latency during microservice deployment in Fog environments. Thus, for Fog computing, it's important to explore distributed container image registries, depending on the resource availability of the Fog infrastructure to host the registry.

\subsubsection{Kubernetes as Container Orchestration Platform}
Decomposition of an application according to microservices architecture results in a large number of microservices and an even more significant number of containers due to horizontally scaled deployment of microservice instances to meet throughput demand, redundant placement of microservice instances to ensure reliability, distributed placement across Fog cluster to support location-awareness, etc. Thus, a container management platform such as Kubernetes is required to manage the life cycle of thousands of containers. 
As one of the most popular open-source container orchestrators, Kubernetes is rapidly improved for use within heterogeneous computing environments through distributions like k3s which is a minimal Kubernetes distribution for extreme edge (i.e., resource-constrained IoT devices, Raspberry Pis, etc.). Thus, the use of k8s and k3s across multi-fog multi-cloud environments is exceeding explored by Cloud providers and Telco providers in their efforts to extend cloud-like services towards network edge \cite{googleEdge, erricsonEdge, ibmEdge}.
Thus, we summarise the basic concepts used in Kubernetes. 
To deploy containers at a scale and to maintain communication among microservice containers, Kubernetes provides build-in \textbf{"resources"} (i.e., Pods, Service, etc.) that provide abstractions for underlying management operations. We discuss some of the most used resources in our framework below.    

\begin{itemize}
    \item Pod: A Pod is the smallest deployable unit supported by Kubernetes, where each pod can contain one or more containers (containers co-located with its sidecar containers). A pod represents a logical host where all co-located containers of the pod share the network resources and communicate through localhost. Pods provide fine-grained control over microservice instance deployment by enabling the deployment of pods on specific nodes by adding node selection constraints (i.e., node selectors, node name, etc.) to the pod.
    \item Service: Kubernetes service is an abstraction over a set of pods within a Kubernetes cluster that provides discovery and load balancing to those pods, thus allowing pods to get dynamically created and destroyed. Although in-cluster service discovery is handled through services, multi-cluster service discovery is not possible with Kubernetes alone. 
    \item Namespace: Namespaces isolate name-spaced Kubernetes objects (i.e., pods, services, etc.), thus providing a way to isolate resources within multi-tenant Kubernetes clusters. 
    \item ConfigMaps: ConfigMaps stores configurations as key-value pairs, thus separating configurations from the pods. This improves the flexibility and portability of containerised microservices.
    \item Secrets: Secrets are similar to ConfigMaps, but are designed to hold sensitive information that should not be stored within the application code.
    \item Roles and Rolebindings: They grant role-based access to Kubernetes resources (i.e., nodes, pods, configmaps, etc.) 
\end{itemize}

\subsubsection{Istio as Service Mesh }
While Kubernetes provides basic functionalities required for container orchestration, it has limitations related to service discovery, load balancing, observability, fault tolerance and security management of the microservice applications. Thus, the service mesh is introduced as a software abstraction layer on top of Kubernetes to overcome these limitations. To this end, Istio implements multiple Custom Resource Definitions (CRDs) extending Kubernetes resource definitions as follows:

\begin{itemize}
    \item Virtual Service (VS): Virtual Services provide more control over traffic routing by providing a way to define traffic routing rules to pods exposed through Kubernetes services.
    \item Destination Rules (DR): Once virtual service routing rules are applied, and the traffic is routed to the destination, Destination Rules are applied to perform load balancing, direct traffic towards service subsets, etc.
    \item Gateway: Gateway is an abstraction for a load-balancer for ingress and egress traffic of the cluster. Furthermore, to support inter-cluster traffic among Kubernetes clusters spread across different networks, Istio provides a specialised gateway known as the east-west gateway.
\end{itemize}

Kubernetes and Istio provide HTTP REST APIs to retrieve, create, update, and delete the above resources. Moreover, client libraries (i.e., Fabric8, client-go, etc.) are available for accessing these APIs through programming languages.

\subsubsection{Example Application Deployment}\label{sed:exampleDeployment}
\begin{figure*}[!ht]
    \includegraphics[width=\linewidth]{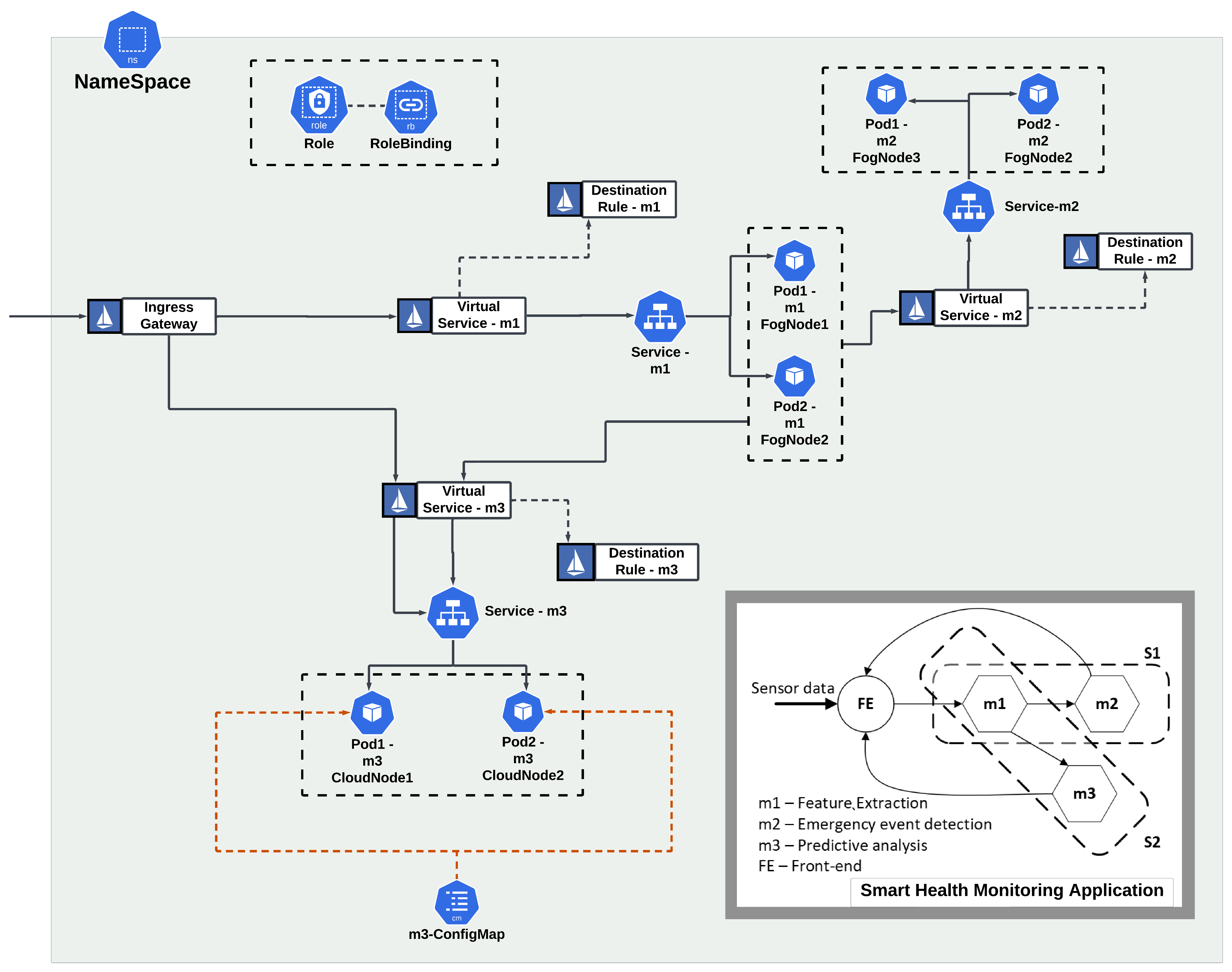}
    \caption{Example Deployment of a Smart Health Monitoring Application}
    \label{fig:exapleAppDeployment}
    \vspace*{-2.0mm}
\end{figure*}

In this section, we demonstrate the use of Kubernetes and Istio resources to deploy a microservices-based IoT application within Kubernetes and Istio available clusters. We use a Smart Health Monitoring Application (see Figure \ref{fig:exapleAppDeployment}) \cite{pallewatta2022qos} as a use case. The application consists of three microservices and two composite services accessed by the users: a latency-sensitive emergency event detection service ($S1$) where both its microservices ($m1,m1$) are placed in distributed Fog resources, a latency-tolerant predictive health warning service consisting two microservices ($m1,m3$). $m1$ is shared between both services and placed within the Fog layer to meet stringent latency requirements of service $S1$, whereas $m3$ is deployed within the Cloud. 

Figure \ref{fig:exapleAppDeployment} demonstrates a logical view of how Kubernetes and Istio resources route external traffic from users to $m1$ and $m3$ and maintain communication between interconnected microservices (between $m1$ and $m2$, between $m1$ and $m3$). 


\subsubsection{Kubernetes + Istio Multi cluster support}
Istio supports deploying a single mesh to span multiple Kubernetes clusters, thus enabling cross-cluster service discovery and load balancing. The Istio deployment model for multi-cluster scenarios depends on the nature of the underlying network model. The simplest network model considers multiple clusters belonging to a single network where all nodes are fully connected through technologies like VPN. However, large-scale production systems that span multiple Kubernetes clusters belong to multiple networks with administrative boundaries where each cluster is exposed through load balancers. Fog computing architecture considered in this work (section \ref{sec:fogmodel}) maps to a multi-network model. Hence, in this work, we consider Istio multi-network deployment with multiple control planes to improve the resilience of the deployment. In this deployment mode, each Istio control plane connects to the API server of the connected clusters for service discovery across clusters. 

 Istio introduces an east-west gateway to expose the services within the cluster to other clusters to enable cross-cluster service discovery. Moreover, to ensure successful DNS lookup across clusters, consumer clusters need to have access to the Kubernetes Service resource, Istion DR and VS of the consumed service deployed in other clusters. As an example, for $S1$ an example application, for $m1$ to route traffic from its Fog cluster to $m2$ deployed within a Cloud cluster, the above resources related to $m3$ should be deployed within both Fog and Cloud clusters.

\subsection{Placement Problem}

Microservice-based IoT application placement problem within Fog environments addresses deployment and maintenance of microservices within federated Fog and Cloud environemnts to meet the Service Level Agreements (SLA) of the application services \cite{guerrero2019evaluation, skarlat2017towards}. 

Due to the flexibility provided by the microservices architecture, placement algorithms aim to incorporate horizontal scalability to meet throughput requirements \cite{guerrero2019evaluation, deng2020optimal,pallewatta2022qos}, location-aware distribution \cite{guo2022joint}, redundant placement to improve reliability \cite{xu2020service}, balanced placement across Fog clusters and Cloud depending on service discovery capabilities \cite{guerrero2019lightweight, pallewatta2022qos}, optimum load balancing and routing \cite{herrera2021optimal}, etc. to efficiently utilise limited Fog resources while satisfying QoS parameters such as makespan, budget, reliability, availability, and throughput.

Execution of placement algorithms can take place as batch placements \cite{samanta2020dyme, pallewatta2022qos} that process multiple application placement requests at once or sequential placements \cite{lera2018availability, guerrero2019lightweight} where queued placement requests are processed one after the other.  Moreover, the placement policies can be developed as centralised \cite{farhat2020reinforcement} or distributed \cite{pallewatta2019microservices} algorithms to achieve placement across distributed Fog and Cloud resources provided by multiple infrastructure providers. 

\subsection{Framework Requirements}\label{sec:requirments}

Based on the background, we summarise the functional and non-functional requirements of a framework for scalable placement of microservices-based IoT applications within federated Fog and Cloud computing environments, as follows:

\begin{itemize}
    \item Multi-fog Multi-cloud microservice placement and deployment: Framework should support execution of placement algorithm across multiple Fog and Cloud clusters using either centralised or distributed operation modes. Accordingly, application microservices need to be deployed by using relevant Kubernetes and Istio resources. 
    \item Seamless microservice composition across hybrid environments: Kubernetes and Istio resource deployment should ensure cross-cluster service discovery and load balancing.
    \item Ability to integrate novel placement algorithms and load balancing policies easily.
    \item Support for heterogeneous cloud-native application deployment without any application-level changes. 
    \item Compatibility with cloud-native technologies so that the framework can improve and evolve as the underlying technologies evolve (extensibility).
    \item A configurable control engine to support different operation modes like centralised or distributed operation, application placement modes such as event-driven or periodic placement request processing and batch or sequential placement request processing.
    \item Distributed storage solutions to store the data required for application placement and deployment (i.e., application models, Kubernetes and Istio resource definitions). 
    \item Rapid prototyping support to enable evaluations of placement algorithms during their rapid design and development cycles.
    \item Framework should be flexible and scalable such that it can be deployed to operate across distributed Fog and Cloud clusters.
\end{itemize}

\subsection{Existing Fog Frameworks}

\setlength\extrarowheight{4pt}
 \begin{table*}[t!]
	\caption{Comparison of existing frameworks}
	\label{table:relatedworks}
	\resizebox{\linewidth}{!}
	{\begin{tabular}[htbp]{ |c|c|c|c|c|c|c|c|c|c|c|c|c|c|c|c|c|
	}
			\hline
			\multicolumn{1}{|c}{\textbf{Work}} &\multicolumn{2}{|c}{\textbf{Architecture}} &\multicolumn{5}{|c|}{\textbf{Cloud-native Application Support}}
                &\multicolumn{5}{c|}{\textbf{Microservice Composition Support}}
			&\multicolumn{3}{c|}{\textbf{Control-engine}}
			&\multicolumn{1}{c|}{}               
			\\
                \cline{2-16}
			Work & Integration & Multi-cluster
			& $\mu$services & Containers & Container  & Service Mesh & Automated  & 
   \multicolumn{2}{c|}{Service Discovery} & \multicolumn{3}{c|}{Load Balancing} &
      Extensibility & Scalability & Configurability & \textbf{Data Stores} 
			\\
                \cline{9-13}
                & & & & & Orchestration & & Deployment & Avail. & cross-cluster & Avail. & configurable & Cross-cluster & & & &  \\
                \hline
                \cite{yousefpour2019fogplan} & Fog, Cloud & - & \checkmark & \checkmark & - & - & $\partial$ & $\partial$ & - & & - & -& \checkmark & $\partial$ & $\partial$ & Centralised \\
                \hline
                \cite{santoro2017foggy} & Fog, Cloud & - & \checkmark & \checkmark & \checkmark & - & $\partial$ & \checkmark (Kubernetes) & - & \checkmark (Kubernetes) & - & - & \checkmark & $\partial$ & $\partial$ & Distributed  \\
                \hline
                \cite{fogatlas} & Fog, Cloud & - & \checkmark & \checkmark & \checkmark & - & $\partial$ & \checkmark(Kubernetes) & - & \checkmark(Kubernetes) & - & - & \checkmark & $\partial$ & $\partial$ & - \\
                \hline
                \cite{ermolenko2021internet} & Edge & - & \checkmark & \checkmark & \checkmark & - & - & \checkmark & - & \checkmark & - & - & \checkmark & $\partial$ & $\partial$ & - \\
                \hline
                \cite{bellavista2017feasibility} & Fog,Cloud & - & \checkmark & \checkmark & \checkmark & - & $\partial$ & \checkmark (Docker Swarm) & - & - & -& -&$\partial$ & $\partial$ & $\partial$ & - \\
                \hline
                \cite{tuli2019fogbus} & Fog, Cloud & - & -& - & - & - & $\partial$ & - & - & - & - & - & $\partial$ & $\partial$ & $\partial$ & Distributed \\
                \hline
                \cite{deng2021fogbus2, wang2022container} & Fog, Cloud & - & $\partial$ & \checkmark & $\partial$ & - & $\partial$ & $\partial$ (Proxy Server) & - & - & -& - & $\partial$ & $\partial$ & $\partial$ & Distributed \\
                \hline
                \cite{mahmud2021pi} & Fog, Cloud & - & \checkmark & \checkmark & - & - & $\partial$ & $\partial$  & - & - & - & - & \checkmark & $\partial$ & $\partial$ & Centralised \\
                \hline
                \textbf{Our} & Fog,Cloud & \checkmark & \checkmark & \checkmark & \checkmark & \checkmark & \checkmark  & \checkmark & \checkmark & \checkmark & \checkmark & \checkmark & \checkmark & \checkmark & \checkmark & Distributed, Replicated \\
                & & & &&&&& (Kubernetes, Istio) &  & (Istio) & & & & & & Fault-tolerant \\
                \hline
			
	\end{tabular}}
	\vspace*{-3.5mm}
        \raggedright\footnotesize{\checkmark: Supported by the framework, $\partial$: Partially supported}
\end{table*}

In this section, we compare existing Fog frameworks qualitatively based on the requirements identified in the previous section (see Table \ref{table:relatedworks}). 

Yousefpour et al. \cite{yousefpour2019fogplan} present a FogPlan, a framework for dynamic provisioning containerised Fog services using container orchestration platforms such as Kubernetes or OpenStack. FogPlan consists of a centralised Fog Service Controller responsible for hosting the data stores, provisioning Fog services and deploying them within Fog nodes. Santoro et al. \cite{santoro2017foggy} provide an open-source technology-based (i.e., OpenStack, Kubernetes, Docker) platform named Foggy for workload placement in Fog computing environments. FogAtlas \cite{fogatlas} extends Foggy platform by extending Kubernetes to orchestrate distributed Fog and Cloud resources in a user-friendly manner. Ermolenko et al. \cite{ermolenko2021internet} also propose a framework based on Kubernetes and Docker where a Kubernetes cluster is deployed within a Mobile Edge Computing (MEC) environment. Bellavista et al. \cite{bellavista2017feasibility} create a microservice deployment framework based on Docker and Docker Swarm with a centralised control engine deployed in the Cloud to execute placement algorithms and deploy microservices accordingly. While they utilise Kubernetes and Docker Swarm features for container orchestration, they also have limitations in multi-cluster support, advanced microservice composition with service mesh technologies, and scalability of the control engine across multi-fog multi-cloud environments. Tuli et al. \cite{tuli2019fogbus} introduced FogBus framework to harness edge/Fog and remote Cloud resources for the placement of applications developed as a collection of inter-connected modules. Deng et al. \cite{deng2021fogbus2} proposed FogBus2, a resource management framework for the deployment of containerised applications across edge and Cloud resources that are interconnected to each other using a VPN network. Wang et al. \cite{wang2022container} improved FogBus2 and integrated container orchestration capabilities to the framework using Kubernetes. Their framework supports the integration of novel placement policies and their performance monitoring to evaluate novel placement policies. However, their framework lacks support for multi-cluster scenarios with multiple geo-distributed Kubernetes clusters. Moreover, they lack support for the dynamic composition of microservices due to limitation in service discovery and load balancing aspects and does not integrate service mesh technologies to fully leverage the capabilities of microservices architecture. Kubernetes resource usage in FogBus2 is limited only to Pods, which limits the framework's scalability. Furthermore, application-level changes are required for the containerised application modules to be deployed within the framework. Mahmud et al. \cite{mahmud2021pi} propose a fully distributed and scalable framework named Con-Pi to execute microservices-based applications. Con-Pi provides a centralised controller to execute integrated customised placement policies and deploy containerised microservices accordingly. However, Con-Pi does not provide advanced microservice composition, dynamic service discovery and load balancing for the deployed microservices and does not consider application deployment across multiple Fog resource clusters. 

Based on the qualitative analysis provided in Table \ref{table:relatedworks}, existing frameworks have limitations in multiple requirements identified in section \ref{sec:requirments} such as multi-fog multi-cloud placement and fully-automated deployment of applications, ensuring cross-cluster dynamic composition of microservices through container orchestrators and service mesh technologies, improving extensible of the framework through open-source technologies, scalability of the framework across highly distributed Fog environments, configurability to support different operation and placement modes, and distributed management of data required for application placement and deployment. Thus, this work introduces a novel framework for microservices-based application placement within federated Fog environments that satisfy the above requirements.

\section{MicroFog Framework}\label{sec:microfog}

In this section, we discuss the high-level architecture of the proposed MicroFog framework, its main components and workflow to highlight how MicroFog meets the requirements identified in section \ref{sec:requirments}.

\subsection{High-level Architecture}

\begin{figure*}[!ht]
    \includegraphics[width=\linewidth]{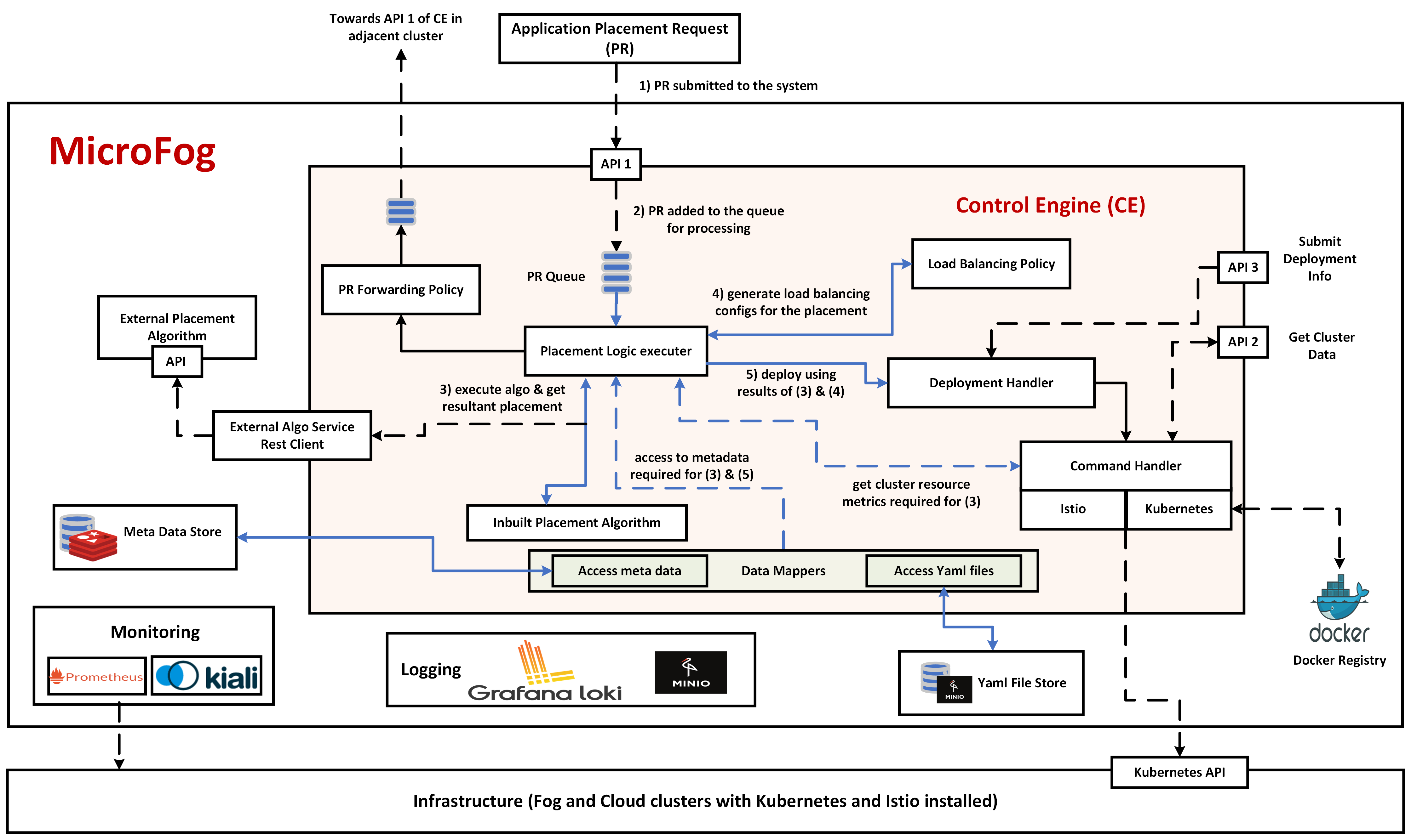}
    \caption{MicroFog: High-level Architecture}
    \label{fig:highlevelarchitecture}
    \vspace*{-3.5mm}
\end{figure*}

Figure \ref{fig:highlevelarchitecture} presents the high-level architecture and the workflow of MicroFog. MicroFog provides a scalable and extensible Control Engine (CE) to execute placement algorithms and deploy IoT applications within Istio-installed Kubernetes clusters. CE communicates with three data stores: 1. YAML File Store containing YAML definitions (both Kubernetes and Istio) required for deployment of applications, 2. Meta Data Store for storing application models and links to related deployment resources stored within the YAML File Store, and 3. Docker registry hosting docker images for the application microservices. Application providers can submit Placement Requests (PRs) to the MicroFog-CE, defining the application for deployment and QoS requirements. CE receives application placement requests (PRs), process them according to a selected placement policy (either an inbuilt placement algorithm or external algorithm accessed through an API), configure related Kubernetes and Istio YAML files according to the generated placement and the load balancing policy, and finally deploy them within Fog and Cloud resources using Kubernetes API. Furthermore, MicroFog integrates monitoring and logging tools to observe the performance of the MicroFog framework and applications deployed using it.

\subsection{Main Components and Technologies }

\subsubsection{Control Engine (CE)}
CE is designed to abstract microservices placement (execution of placement algorithms and deployment) and cross-cutting function handling (i.e., service discovery, load balancing) for the dynamic composition of microservices across multi-fog multi-cloud environments.

We implement CE as an independently deployable and scalable microservice developed using Quarkus \footnote{https://quarkus.io/}, a novel Kubernetes-native lightweight Java framework designed to build cloud-native microservices. Quarkus reduces memory usage and improves deployment density \cite{quarkusVal}, which is suitable for developing microservices for deployment within resource-constrained Fog environments. As Quarkus is a Kubernetes-native framework, the development and deployment of the CE become straightforward and less time-consuming, thus allowing users to rapidly improve, extend and customise it with evolving needs. Thus, Quarkus is rapidly becoming popular as a lightweight Java framework for creating cloud-native microservices. Moreover, Quarkus allows easy access to Fabric8 Kubernetes and Istio clients \footnote{https://github.com/fabric8io/kubernetes-client} through its extensions. Fabric8 is a highly popular Kubernetes and Istio client that provides complete access to Kubernetes API. Fabric8 consists of a rich DSL (Domain Specific Language) for interacting with Kubernetes API, hence making it one of the most used open-source Kubernetes clients with an extremely active community using and continuously improving it. Thus, we have selected Qaurkus together with Fabric8 Kubernetes and Istio clients to create our controller. 

We discuss the functional and non-functional features of the MicroFog-CE as follows:

\begin{enumerate}[leftmargin=*]
    \item \textit{PR submission for placement: } Application providers can submit their PRs to the CE through an API which expects HTTP POST requests with the PRs represented in JSON format (API 1 shown in Figure \ref{fig:highlevelarchitecture}). Each submitted PR can define multiple data fields related to the application, including application id, QoS parameters, any restrictions for application placement, traffic entry clusters, etc. Once submitted, CE uses such information to process the PR (i.e., the application id is the key to retrieving the application model and deployment resources from the data store, and entry clusters denote the clusters that act as the entry point for the ingress traffic for the considered application) and deploy the application microservices and deployment resources accordingly.

    \item\textit{Multiple operation and placement modes:} CE supports Centralised and Distributed operation modes. In centralised mode, a primary CE (i.e., deployed within the Cloud) with a global view of the infrastructure (i.e., Fog, Cloud clusters, their topology and resource availability) is responsible for executing the placement algorithm. In this mode, the primary CE queries the secondary CEs (through API 3) to gain information regarding the resources available within each cluster and their topology-related data (i.e., directly reachable Fog and Cloud clusters from each cluster) to construct the global view of the federated environment. Primary CE uses this information to generate placements for the applications requested by the PRs and send the output placement details to each relevant cluster (through API 4). The secondary CEs deployed within each cluster process the placement output and deploy Kubernetes and Istio resources accordingly. In contrast, in the distributed mode, all CEs are responsible for running the placement algorithm locally per cluster. They collaborate by forwarding the PRs among the clusters for distributed placement across multi-fog multi-cloud environments. MicroFog-CEs use API 1 for PR forwarding among clusters as well.

    Furthermore, the CE supports two placement modes: Periodic Placement and Event-driven Placement. Periodic placement invokes the placement algorithm periodically based on a configurable time period. Under this mode, the placement algorithms can be designed to process the PRs either as a batch (all PRs in the queue are processed simultaneously by the algorithm) or sequentially (either in First-In-First-Out order or prioritised). In the event-driven mode, the placement algorithm is invoked upon receiving a new PR.

    \item \textit{Placement Algorithm Integration:} CE supports easy integration of novel placement algorithms. This can be done using two methods: in-built algorithm implementation where novel placement policies can be implemented by extending \textit{PlacementAlgorithm.java} base class of the CE. The base class is initialised with the metadata required by the placement algorithms (i.e., resource availability of the devices, application model and topological information). Novel placement algorithms can extend this to implement customised placement logic that utilises the metadata to produce placement output (denoted by PlacementOutput.java) consisting of microservice-to-device mapping and PR completion data (completed PRs vs incomplete PRs that should go through a forwarding process to other clusters for placement completion). Moreover, CE provides capability to integrate external placement algorithms, which allows algorithms to be implemented in other programming languages (i.e., Python for placement algorithms that use Machine Learning). Such algorithms can be implemented as a separate microservice and integrate it to the MicroFog-CE by implementing an API that can be called by the \textit{External Algo Service Rest Client in Figure \ref{fig:highlevelarchitecture}} of the CE through an HTTP GET request. CE rest client is designed to send the metadata along with the GET request so that the external placement algorithm can generate the placement and return the deployment-related information back to the CE.

    By default, MicroFog-CE implements a Latency-aware Scalable Placement Policy proposed in \cite{pallewatta2019microservices}. The above algorithm aims to place microservices of latency-critical service as close as possible to the users who access them. We implement this algorithm in both distributed and centralised modes. We also implement it with and without horizontal scalability of the microservices to demonstrate the performance improvement MSA can provide within resource-limited Fog environments. 

    \item\textit{Access Infrastructure Metrics: } To make placement decisions, placement algorithms require metrics related to infrastructure, such as resource availability within the cluster. To this end, the current version of CE provides two measurements: 1. CE access Kubernetes Metric Server to obtain node metrics of current CPU and RAM usage, 2. CE also provides current resource allocation of the deployed pods by querying the Kubernetes API. Placement algorithms can utilise both types of metric information to make placement decisions. Metric collection can be further extended to use Prometheus as well to utilise time-series metric data for placement decision making. 

    \item \textit{Load Balancing Policy Integration: } Due to the independently deployable and scalable nature of the microservices, load balancing plays a vital role in properly distributing the load across horizontally scaled microservices deployed across federated Fog and Cloud environments. By default, Istio use a round-robin load balancing method to route the requests. Moreover, Istio supports other load balancing methods like random, least request and weighted load balancing, which are already implemented in Envoy Proxy used by Istio for service discovery and load balancing purposes. They can be configured by updating the Istio DRs related to each microservice. In addition to thus, MicroFog-CE provides enhanced capabilities to support custom load-balancing policies, where weights of the weighted load-balancing approach can be updated based on custom load-balancing policies. 
    
    As an example, the current version of the CE implements weighted round-robin load balancing policy. Once the weight for each microservice instance is calculated based on the placement, CE handles the updates related to subsets, weights, and routes in Istio VS and DR resources. While this update is straightforward for centralised operation mode, distribute placement has one main challenge. Load balancing information can only be calculated after all required microservice instances are placed. Moreover, to execute load-balancing policies properly, Istio needs VS and DR resources to be available in all clusters that host the particular microservice (consumed microservice) and any microservice that tries to interact with it (consumer microservices). Thus, in distributed placement mode, for each microservice, the CE waits until all its instances and its consumer microservices are placed. Afterwards, the information required for VS and DR updates (subset names and weights) are sent to relevant clusters through API 3 of the distributed CEs.

    \item\textit{PR Forwarding Policy Integration:} Placement across multi-cloud multi-fog environments requires the use of distributed placement policies across infrastructure provided by multiple Cloud and Fog infrastructure providers. MicroFog-CE enables this by providing the ability to update the status of the partially processed PRs and forward them to adjacent Fog or Cloud clusters. Such PRs are submitted to the selected cluster's API 1. Moreover, novel forwarding policies can be integrated as well. The default implementation of the CE provides two forwarding policies where the PRs can be either forwarded to a random Fog cluster or to the Cloud. As CE instances are configured independently, it is possible to use different forwarding policies across clusters.

    \item \textit{Automated Application Deployment:} MicroFog CE abstracts the microservice deployment process from the framework users. For each application, YAML File Store is used to retrieve the Kubernetes and Istio resources related to the deployment of microservices. This includes resources at different abstraction levels such as 1. application level resources such as Namespaces, Roles and RoleBindings, 2. microservice level resources such as ConfigMaps, Secrets and Pod definition YAML files to create microservice instances on mapped nodes based on the placement algorithm output, 3. Services, Virtual Services and Destination Rules for service discovery across clusters and to load balance and route traffic to create composite services based on the load balancing policy and 4. Gateways to enable ingress traffic to reach root microservices of application DAG. Moreover, MicroFog-CE enables federation across multiple Fog and Cloud clusters by deploying microservice composition-related resources (i.e., Kubernetes Services, Virtual Services, Destination Rules) in relevant clusters. CE rules are designed to handle these functionalities, thus abstracting the underlying complexities from the framework users.

    \item \textit{Scalable and Distributed CE deployment: } As the CE is developed as a microservice using a Kubernetes-native microservice framework, it can be deployed within Kubernetes and Istio-enabled environments in a distributed manner. Each CE can be configured separately and communicate across clusters using the REST APIs, thus making MicroFog scalable to operate across federated Fog and Cloud environments.

    \item \textit{Extensibility:} Design and architecture of the CE capture the problem domain of microservices-based application placement by implementing java objects as rich domain-specific objects. Figure. \ref{fig:domainDiagram} domain diagram used in developing the MicroFog-CE, which adheres with the system models and placement problem formulated in the section \ref{sec:background}. This makes the CE implementation easy to comprehend and extend to incorporate novel features. Moreover, due to the compatibility of the MicroFog framework with open-source cloud-native technologies, the CE can evolve as the capabilities of the underlying technologies evolve. 

    \item \textit{Configurability:} Quarkus enables application configuration properties to be acquired through Kubernetes ConfigMaps. This highly improves the configurability of the CE, where the users can update application configurations without creating new Docker images to rapidly use different configurations (policies, placement modes, operation modes, etc).
\end{enumerate}

\subsubsection{Data Stores}

MicroFog uses three main data stores as follows: 
\begin{enumerate}[leftmargin=*]
\item\textit{Meta Data Store:} Metadata store contains application-related information belonging to two main categories: 1) application model (as discussed in section \ref{sec:appModel}) which contains specification related to microservices, interconnections among microservices to create services, dataflows, etc. 2) application deployment related Kubernetes and Istio resources. This includes resource type (i.e., Namespaces, Pods, Services, etc.) and URL to the YAML file containing the specifications of each resource. We use Redis \footnote{https://redis.io/} as a primary database to store this information. Even though Redis was initially introduced as a cache, now it is increasingly used as a primary database to reduce the complexity of data retrieval and improve performance. Redis allows data to be stored as key-value pairs. With the use of Redisson, a Redis Java client, the $Application$ domain objects of the CE can be easily serialised to store within the Redis metadata store and retrieve them back as Java objects. 

\item\textit{Yaml File Store: }This is used for storing Kubernetes and Istio resource configurations as YAML files. Due to the geo-distributed nature of the Fog clusters, a distributed object store is required for efficiently storing the YAML files. To meet this requirement, we use MinIO Object Store \footnote{https://min.io/}, an AWS S3 compatible, Kubernetes-native object store designed for multi-fog multi-cloud environments. For each Istion/Kubernetes resource to deploy, the CE retrieves the YAML file from the MinIO data store using an object URL and uses the Fabric8 Kubernetes client library to load it as a domain object representing the deployment resource. 

\item\textit{Docker Registry: } As IoT application microservices are containerised for deployment, the container images must be stored in a docker registry reachable by the CEs. In the current implementation, we use Docker Hub, a publicly available managed Docker store. However, this can be further improved by using local Docker stores in conjunction with Docker Hub, depending on the resource availability of each Fog cluster to host the images. 
\end{enumerate}

\subsubsection{Monitoring and Log Management: } Due to their highly distributed and dynamic nature, monitoring and observability remain essential aspects of cloud-native microservices. To this end, Istio enables the integration of multiple tools in the form of pre-configured plugins. This includes metric collection and visualisation (Prometheus and Grafana), distributed tracing (Jaeger, Zipkin), and mesh visualisation using Kiali. In the current version of the MicroFog framework, we have integrated Prometheus, Kiali and Grafana to observe the traffic across clusters and to validate the functionalities of the MicroFog-CE. In addition, MicroFog uses a cluster-level logging architecture to manage the logs generated within each cluster. To this end, MicroFog uses Grafana Loki, a decentralised, lightweight logging stack that compresses and stores data in object stores such as S3. As the MinIO object store used for YAML File storage is S3 compatible, MicroFog uses the same store for storing the logs. Compared to other cloud-native logging solutions like ElasticSearch, Loki has a less complex architecture, requires less storage and consumes less power, which makes it suitable for Fog deployment. Depending on the resource availability of the Fog clusters, the logs can be stored within the MinIO hosted in Cloud to save storage space. However, other tools also can be easily integrated depending on requirements. Moreover, the current architecture can be easily extended so that MicroFog-CE can use the metrics collected from monitoring and logging tools to execute dynamic placement algorithms or integrate machine-learning-based approaches.

\subsubsection{Rapid Prototyping Support}
Producing novel placement algorithms undergo multiple development and evaluation cycles to optimise their performance. Thus, rapid prototyping during different stages of policy development is beneficial before conducting large-scale evaluations or applying them in real-word application deployments. Due to the use of open-source cloud-native tools, MicroFog enables fast creation of underlying infrastructure using tools such as Kind and MetaLB to create Fog computing clusters consisting of heterogeneous nodes and route inter-cluster traffic through load balancers. 

\subsection{PR Processing flow of MicroFog-CE}
In this section, we discuss the high-level pseudo-code (see Algorithm \ref{alg-ce}) of the MicroFog-CE with regards to processing received PRs. In an environment where each cluster contains a separate CE, the depicted PR processing procedure is executed in all CEs under the distributed placement mode and only in the primary CE if the placement mode is set to centralised placement.

PR processing begins with retrieving \textit{PR}s from the \textit{PRQueue} (line 1). The method of retrieval depends on the placement mode of the CE, where in periodic placement, all PRs collected in the \textit{PRQueue} are retrieved for processing, whereas in event-driven mode, each PR is taken from the queue as its added. If the PR processing thread is busy, the PR waits in the queue until the thread becomes free. The current implementation of the CE uses a single thread for the PR processing, whereas multiple threads add incoming requests to the \textit{PRQueue} implemented using \textit{Java ConcurrentLinkedQueue}, which is a non-blocking and thread-safe queue implementation.

\begin{figure*}[!ht]
    \centering
    \includegraphics[width=0.8\linewidth]{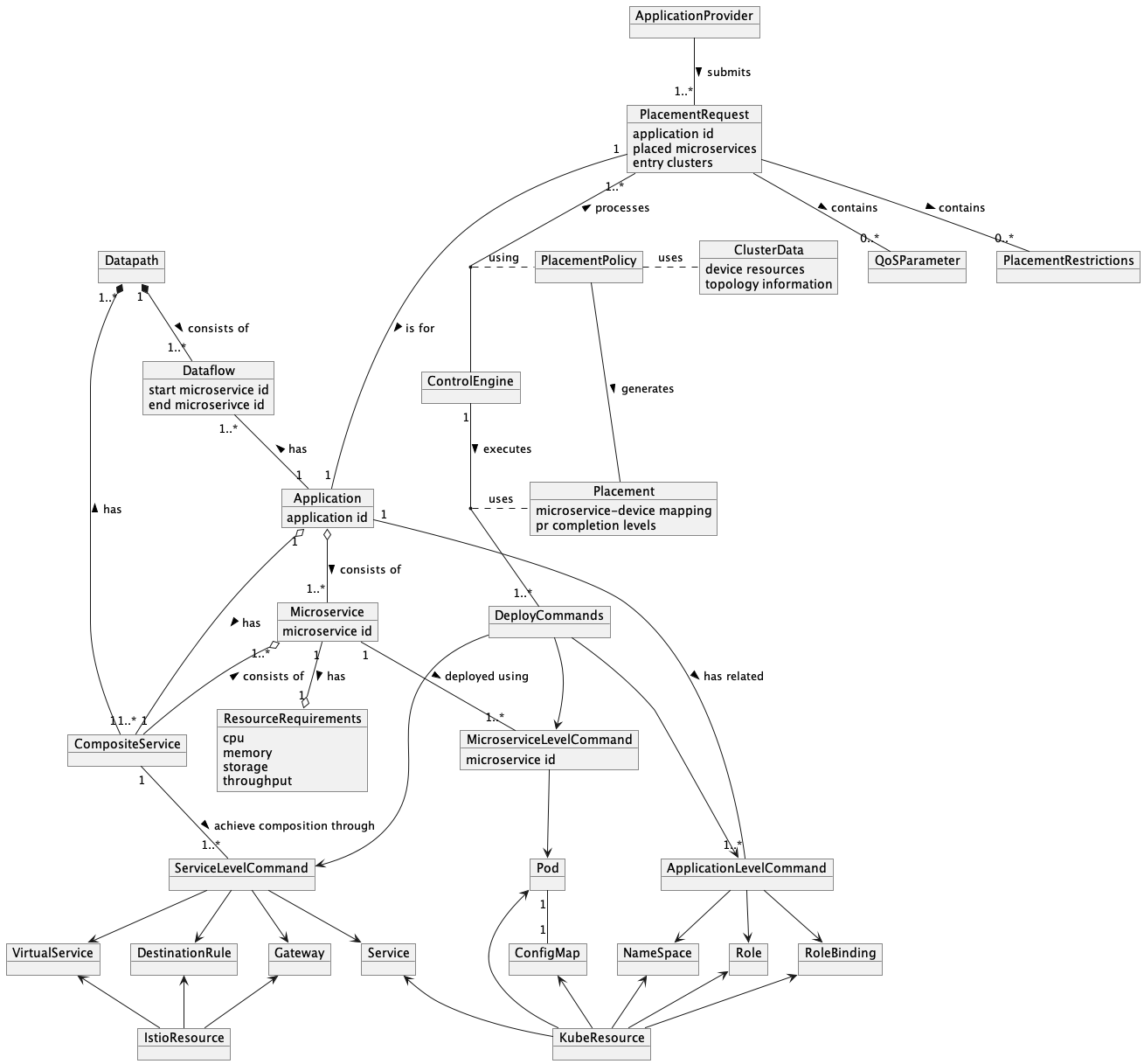}
    \caption{MicroFog: Domain Diagram for Control Engine}
    \label{fig:domainDiagram}
    \vspace*{-2.0mm}
\end{figure*}

Retrieved PRs undergo three main steps: Meta Data Retrieval, Placement Algorithm Execution, and finally, Deploying microservices-based applications using Kubernetes and Istio resources and handling uncompleted PRs. The first step of metadata retrieval is to generate cluster data required by the placement algorithm (lines 5-11). This includes details about the resource availability of each node in the cluster along with topological details such as adjacent Fog and Cloud clusters of each considered cluster. For centralised placement, the primary CE that is responsible for executing the placement algorithm needs to have a bird's eye view of all the Fog and Cloud clusters. Thus, the primary CE queries other clusters by sending requests to the API 2 of the connected clusters (lines 10-11). For this, we implement a Reactive REST Client that sends all requests simultaneously, waits for the results of all the sent requests, and retrieve each cluster's data from the reply. Reactive REST Clients supported by the Quarkus framework enable concurrent request sending, which improves the efficiency of collecting data from distributed clusters. As the second step of metadata retrieval, the CE queries the application model related to the application requested by each PR from the Redis metadata store (line 13). This retrieves a Java domain object of type $Application$ (as depicted in domain model \ref{fig:domainDiagram}) which consists of Microservices, Composite Services, Datapaths, Dataflows, Resource Requirements and Commands used for microservice deployment, which are all depicted using serialisable Java objects. 

Afterwards, the CE starts processing the \textit{PR}s using the placement algorithm (lines 16-19). As the CE can support integration of placement algorithms either by extending the existing CE or as an external microservice, the algorithm can be configured as a property of the CE. The CE is designed to use the factory pattern to initialise placement algorithms based on the configured placement algorithm name. Thus, the internal integration of the placement algorithms requires them to be added to the factory. To use external algorithms, CE implements a REST client with a configurable URL that can be updated with the URL of the external algorithm (line 19). 

Once the placement output is generated by the placement algorithm, the CE moves on to the application deployment stage. During this step, CE generates deployment information for each cluster under two main categories: basic deployment information and load balancing information. Basic deployment information includes pod-to-device mapping with required resource allocation, ingress clusters for each application for the deployment of Istio Gateway and related Virtual Service for ingress traffic routing, etc. Load balancing-related deployment information generation includes executing the load balancing policy for the placement of completed microservices and generating subsets and weights accordingly. This data will be used to update Virtual Services and Destination Rules to ensure desired load balancing. 

After generating the deployment information, the CE invokes a new thread to forward incomplete PRs (in the distributed placement mode) based on the forwarding policy while the current thread continues with deployment. In the centralised placement mode, the CE uses a Reactive REST Client to send the deployment information to others concurrently while the deployment for the current cluster is carried out in parallel as well. This decision is made to improve the overall efficiency of the placement as the deployment of microservices as Docker containers can be time-consuming if carried out sequentially. Similarly, in the distributed placement mode, load balancing information relevant to previous clusters are also transmitted concurrently while one thread continues with deployments related to the current cluster.

\begin{algorithm*}[!t]
		\footnotesize
		\caption{MicroFog-CE PR processing}\label{alg-ce}
		\begin{algorithmic}[1]
			\Procedure{ProcessPRs}{\textit{PRQueue}}
			\State \textit{PRs} $\gets$ get \textit{PRs} from the \textit{PRQueue} for processing
                \State {\textbf{\# Step 1. Meta Data Retrieval wich consists of two sub-steps 1.1 and 1.2}} 
                \State{\textbf{\# Step 1.1 : Cluster data retrieval (including both resource availability within cluster and tipology information }} 
			\State \textit{clusterData} $\gets \{\}$ \Comment{Maps cluster name to its data}
			\State \textit{inclusterDeviceData} $\gets$ loadInClusterDeviceData()  \Comment{Device data related to the current cluster is loaded}
			\State $currentClusterData \gets inclusterDeviceData \cup topologyData$
			\State $clusterData.$add$(currentClusterName, currentClusterData)$ 
                \State{\textbf{\# For centralised placement, request cluster data from other cluster using API 2}}
                \If{centralisedPlacement \textbf{AND} is primary CE}
                    \State $clusterData  \gets$ requestOtherClusterData()
                \EndIf 
                \State{\textbf{\# Step 1.2 : Loading application meta data form the Meta Data Store }}
                \State \textit{appInfo} $\gets$ loadRelatedAppInfo(prs) 
                \State{\textbf{\# Step 2: Execute the placement algorithm        }}
			\State \textit{placementOutPut} $\gets \{\}$
                \If{is internalAlgo}
                \State $placementOutPut \gets placementAlgo.$generatePlacement($PRs, appInfo, clusterData)$ 
                \EndIf
                \If  {is externalAlgo}
                \State \textit{placementOutPut}  $\gets externalPlacementAlo.$generatePlacement($PRs, appInfo, clusterData, externalUrl)$
                \EndIf
                \State {\textbf{\# Step 3. Deploy using Istio + Kubernetes resources and handle incomplete PRs}} 
                \State \textit{perClusterDeploymentInfo} $ \gets \{\}$
                \State $perClusterDeploymentInfo.$add(generateBasicDeploymentInfo($placementOutPut))$
                \State $perClusterDeploymentInfo.$add(generateLoadBalancingRelatedDeploymentInfo($placementOutPut))$
                \If{is distributedPlacement}  \Comment{Uses a separate thread}
                \State $incompletePRs \gets placementOutPut.$getIncompletePRs()
                \State forwardIncompletePRs($incompletePRs$)
                \EndIf
                \State $thisClusterDeployment \gets perClusterDeploymentInfo.$getThisCluster()
                \State $deploymentHandler.$deploycommands($thisClusterDeployment$)
                \State sendToOtherClusters($perClusterDeploymentInfo-thisClusterDeployment$)
			\EndProcedure
		\end{algorithmic}
	\end{algorithm*}

\section{MicroFog  Deployment} \label{sec:deployment}

Deployment of MicroFog within federated fog-cloud environments includes two main steps: 1. distributed setup for data stores, and 2. distributed deployment of the CE. As example deployment scenarios, we provide deployment architecture (Figure \ref{fig:yamlStore} and Figure \ref{fig:controlengine} for each step. The demonstrated examples consider a federated fog-cloud environment consisting of two Fog clusters and one Cloud cluster. Three clusters belong to three separate networks and are three independent Kubernetes clusters interconnected through Istio multi-primary architecture to enable inter-cluster microservice composition and traffic.

\subsection{MinIO YAML File Store Deployment}\label{Minio}
\begin{figure*}[!ht]
    \includegraphics[width=\linewidth]{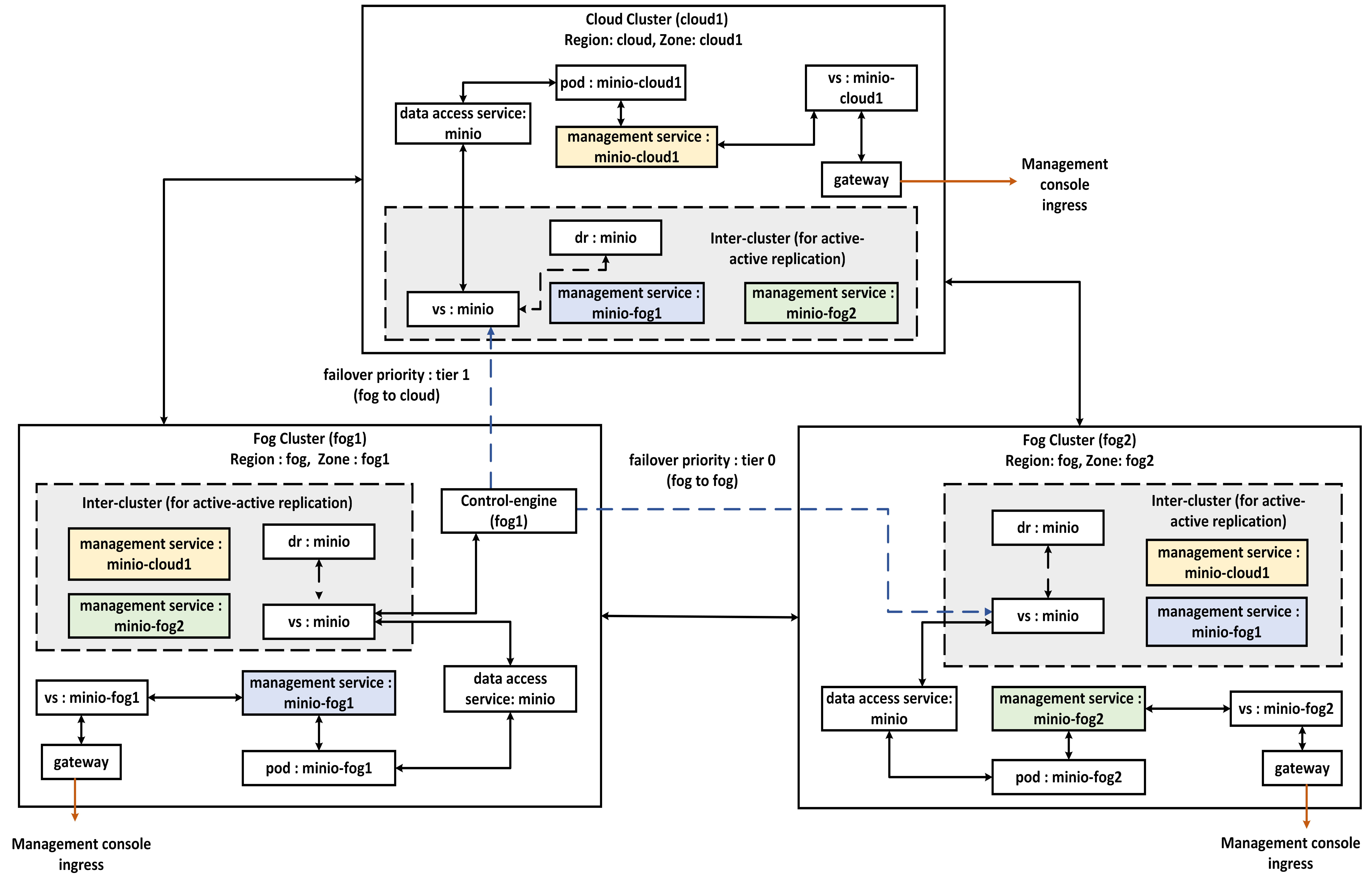}
    \vspace*{-7.0mm}
    \caption{MinIO - YAML File Store - Deployment}
    \label{fig:yamlStore}
    \vspace*{-3.0mm}
\end{figure*}

We provide an example deployment scenario in Figure \ref{fig:yamlStore} to demonstrate the distributed deployment of the MinIO YAML File Store within federated fog-cloud environments. For distributed storage and access of YAML files, we design the deployment architecture to meet the following requirements: 1) Distributed deployment across clusters to improve the latency of application deployment, 2) Replication across distributed data stores to maintain data consistency, 3) Fault-tolerance through a prioritised failover mechanism to ensure availability in a latency-aware manner.

To achieve these objectives, we create two traffic routing layers using Kubernetes and Istio resources, namely, the Management layer and the Data Access layer. The management layer is used for configuring individual MinIO servers deployed per cluster. Kubernetes service and Istio VS for the management layer expose default MinIO ports for management console access through ingress gateway (console port) and data replication among distributed MinIO instances (API port). The second layer of routing exposes the API port of the MinIO data store, for access by the CE to retrieve YAML files required for application deployment. This layer of traffic implements a two-tier failover policy to improve the reliability of the deployment. Istio supports locality-aware load-balancing to failover based on region ( topology.kubernetes.io/region), zone (topology.kubernetes.io/zone) and sub-zone (topology.istio.io/subzone) of the nodes. We use the region and zone to conduct the failover where all Fog level resources belong to the region "fog", where each Fog cluster is considered as a separate zone. Similarly, all Cloud clusters belong to the region "cloud". Istio default failover policy assigns high priority to failover within the same region (i.e., Fog clusters would fail over to adjacent Fog clusters). We further extend this by incorporating an Istio DR to ensure failover from Fog to Cloud if no Fog clusters are available. To ensure proper fault tolerance, each node in the Kubernetes clusters needs to be annotated with their related region and zone. Although the number of tiers is limited to two in the current implementation, it's possible to extend it to three tiers by implementing Istio sub-zones as well.

\subsection{Redis Meta Data Store Deployment}
Deployment of Redis Meta Data flow follows a similar approach with two traffic layers, one for data replication and the other for retrieving application information. We use the master-replica deployment supported by Redis. In our proposed architecture, we deploy the master Redis server in the Cloud cluster and deploy the rest as replicas where they sync with the master server to retrieve the available metadata. Similar to MinIO YAML Store, this deployment also uses locality load-balancing in Istio to ensure failover from the Fog layer to the Cloud to improve the availability of the data.

\subsection{Control-Engine Deployment}\label{sec:ceDeployment}

\begin{figure*}[!ht]
    \includegraphics[width=\linewidth]{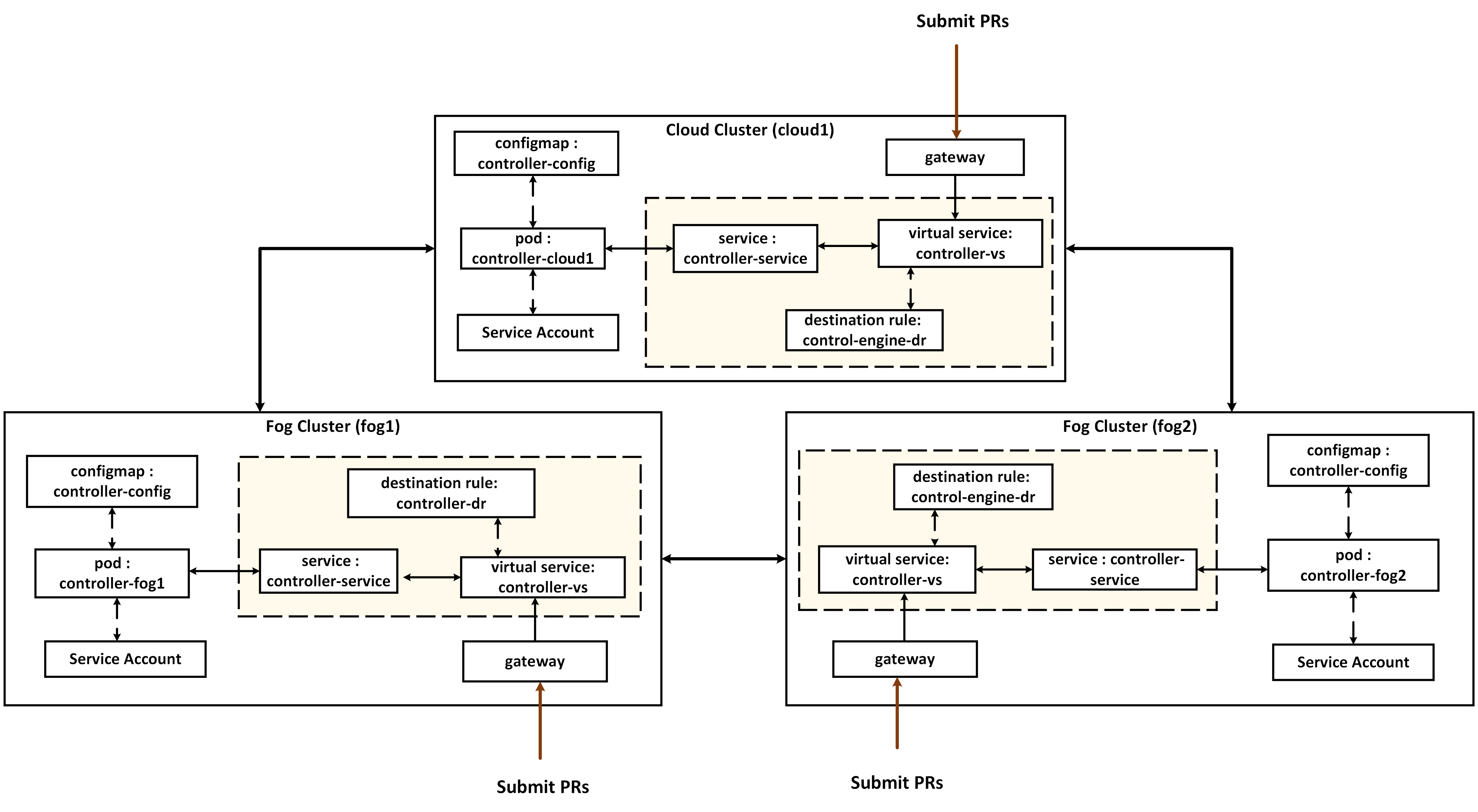}
    \vspace*{-5.0mm}
    \caption{Distributed Control Engine Deployment}
    \label{fig:controlengine}
    \vspace*{-3.0mm}
\end{figure*}

Figure \ref{fig:controlengine} depicts an example scenario for the distributed deployment of CEs across federated Fog and Cloud clusters. We discuss the main aspects of the deployment as follows:
\begin{itemize}
    \item Distributed deployment of CEs and maintaining communication across clusters: In both centralised and decentralised placement modes, CEs need to access APIs of the other CEs deployed in different clusters for various functions, including querying cluster data, forwarding PRs, submitting deployment information. We enable this by using Istio DR and VS to route based on the header value of each request. We introduce a header called "cluster", which defines the destination cluster to route the requests. To achieve proper routing, each pod of CE is labelled with its cluster name, and the DR creates subsets based on the cluster name. Following this implementation, the VS routes by matching the header value to the subset label.
    \item  PR submission to a particular cluster: The above implementation enables not only inter-CE routing but enables ingress traffic to the CE (i.e., submitting PRs) to be routed to a specific CE based on the header value. 
    \item Configure each CE separately during deployment: To improve the efficiency of configuring the CEs and to enable each CE to be configured independently, we use a Kubernetes CongfigMap to define the CE configurations. Due to its Kuenernetes-native nature, the Quarkus application is configured to retrieve the values for application.properties from the ConfigMap.
    \item Ensure access to underlying Kubernetes and Istio deployments: CE needs to access Kubernetes API for various actions (i.e., retrieve node data, retrieve resource metrics, retrieve pod data, deploy Kubernetes and Istio resources). To this end, the proper level of permission should be granted to the CE. A dedicated service account is created and attached to a ClusterRoleBinding and a ClusterRole to grant the required access across the cluster.
\end{itemize}

\subsection{Deployment of Observability, Monitoring and Logging Tools}
 For the current implementation, we integrate Prometheus and Kiali to verify the feature supported by the CE. Kiali uses the Prometheus monitoring tool to create topology graphs, calculate health and show metrics. Istio add-on preconfigures it to visualise multi-cluster service mesh, including different views such as graphs (depicting application, services, microservice versions, etc.), traffic flows, metric details, and Istio configurations (YAML files related to each deployed Istio resource). Within the distributed architecture, Prometheus and Kiali components are deployed per cluster, and the Kiali dashboard is exposed through the Istio ingress gateway to access it remotely. 

 For log aggregation and visualisation we use Loki and Grafana. Loki is configured to use a object bucket from MinIO object store. As the MinIO deployment and request routing is already handled (section \ref{Minio}), logs can be directed either to a central Cloud or stored within the own cluster depending on the resource availability.

\section{APIs of MicroFog-CE}\label{sec:apis}
\begin{figure}[!ht]
    \includegraphics[width=\linewidth]{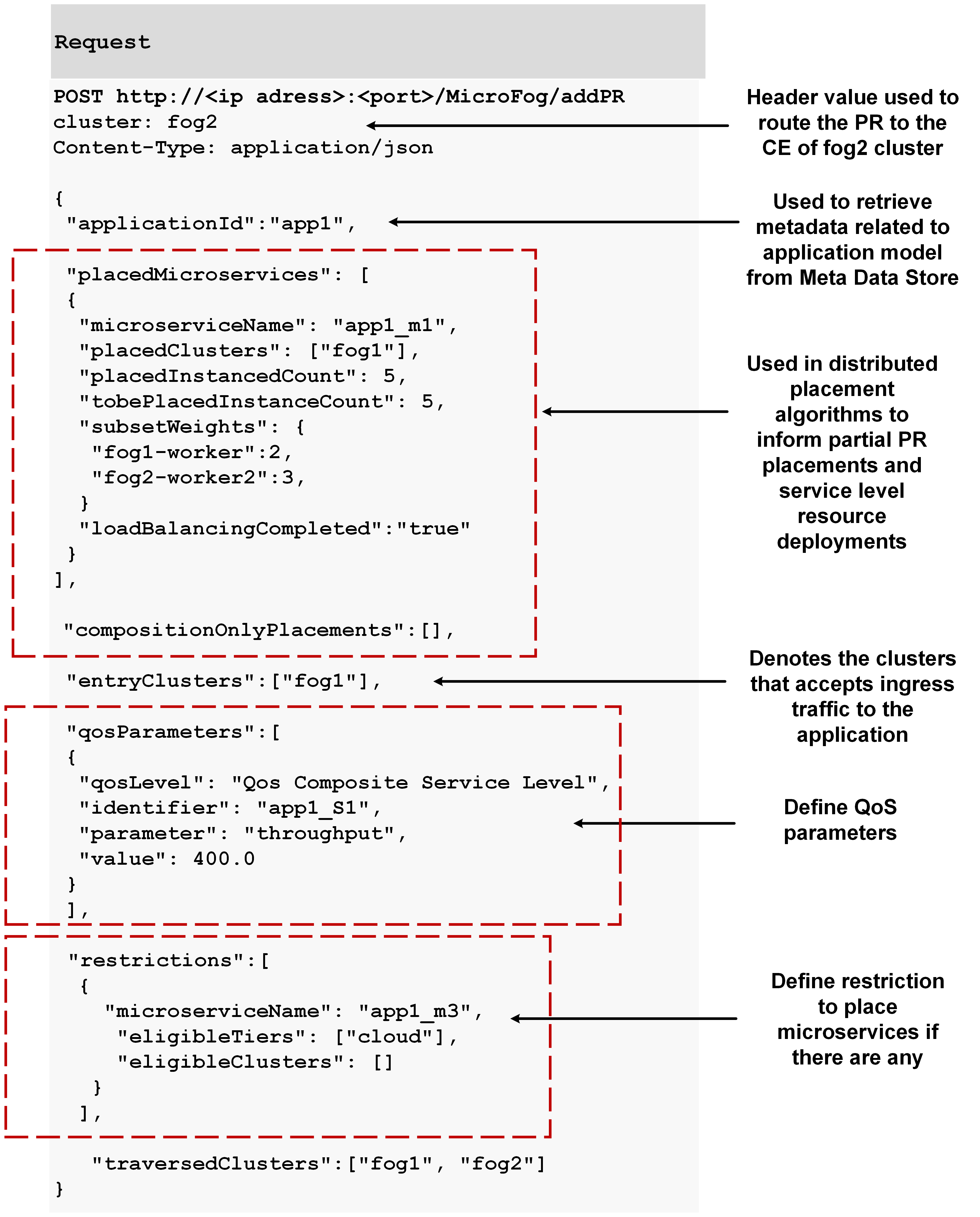}
    \caption{API 1 - For submitting PRs}
    \label{fig:api1}
    \vspace*{-2.0mm}
\end{figure}

\begin{figure}[!ht]
    \includegraphics[width=0.85\linewidth]{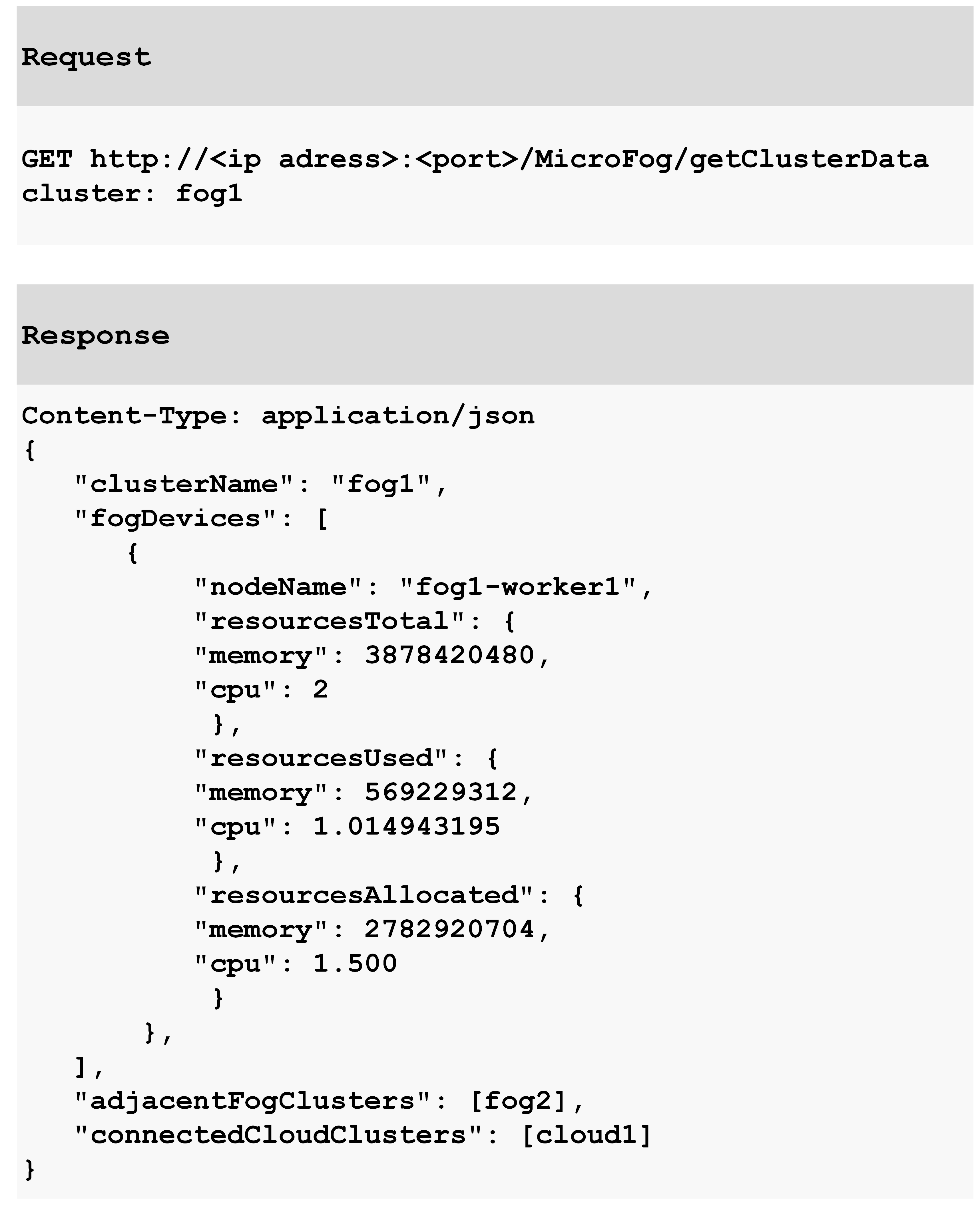}
    \caption{API 2 - For querying cluster information}
    \label{fig:api2}
    \vspace*{-2.0mm}
\end{figure}

\begin{figure}[!ht]
    \includegraphics[width=\linewidth]{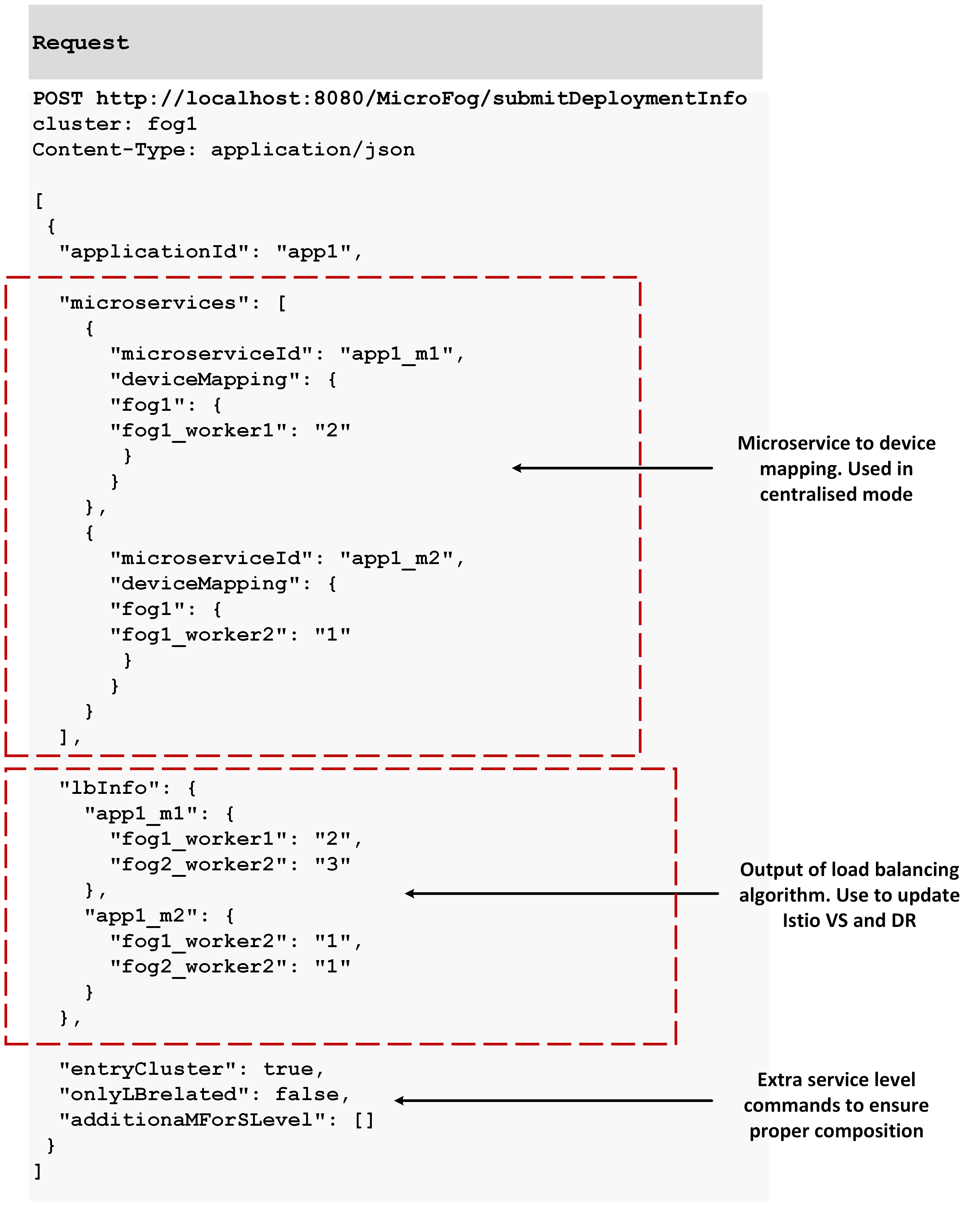}
    \caption{API 3 - For submitting placement output for deployment}
    \label{fig:api3}
    \vspace*{-2.0mm}
\end{figure}

In this section, we highlight the three main APIs provided by MicroFog-CE and also explain the API implementation required to integrate external algorithms into the CE.
\begin{itemize}
    \item API 1 (see Figure \ref{fig:api1}): API 1 is designed for receiving PRs through POST requests, where the request is routed to the cluster defined in the header. The request contains data related to the PR in JSON format, which will be mapped into a Java-based domain object by using the Jackson framework upon receipt. \textit{"applicationId"}, which is used to identify the application to be deployed (matched with the metadata available in the Redis Meta Data Store), and the \textit{"entryClusters"}, which indicates the traffic entry points to the application are required fields for the request data whereas other fields are optional. The rest of the fields are optional and can be filled if relevant. \textit{"placedMicroservices"} indicate already placed microservices and their status. Thus this is mostly used for forwarding requests and can also be used for initial PR submission if some of the application microservices are excluded for placement within Fog or Cloud (i.e., already placed within IoT devices or client devices). 
    \textit{"compositionOnlyPlacements} keep track of intermediate clusters that needs to host service level resources to enable compositing of microservices across non-adjacent clusters. Boolean for \textit{"loadBalancingCompleted"} indicates if load balancing-related deployment information for the microservice has already been transmitted to relevant clusters, whereas \textit{"subsetWeights"} indicate relative resource-allocation among devices to be used for executing load balancing policy. Due to complex dependencies among microservices, the QoS parameters can be defined at multiple granularity levels: per composite service, among microservices and per application \cite{pallewatta2022microservices}. \textit{"qosParameters"} field allows detailed parameter definitions adhering to this. 

    \item API 2 (see Figure \ref{fig:api2}): API 2 is used in centralised placement mode for querying cluster data from each cluster by defining the cluster name in the header to ensure routing. The response returns two main types of data: 1) an array containing resource availability of each node in the cluster defining total resources, resource usage at the time of query and allocated resources (i.e., memory in bytes and CPU in the number of cores/ vCPUs), 2) data related to topology containing the names of adjacent Fog and Cloud clusters.

    \item API 3 (see Figure \ref{fig:api3}): API 3 is for transmitting deployment information to each cluster specified by the header field. For centralised mode, this includes both microservice deployment and load balancing related Istio resource deployment, whereas, in distributed mode, it is limited to load balancing related resources. This API also accepts some additional information, such as the Boolean indication if the cluster is the entry cluster for the application so that the Istio Gateway and VS resources can be deployed accordingly to enable ingress traffic to reach the application. The request also contains a file that includes a list of microservices (\textit{additionalMForSLevel)}, where their service level resources (i.e., Kubernetes Service, Istio VS and DR) need to be deployed within the cluster to maintain seamless connectivity among microservices deployed within clusters that are not adjacent.
\end{itemize}

Due to the use of Jackson library for conversion between JSON data and JAVA domain objects, the data sent to/from APIs can be modified easily by updating the relevant domain objects accordingly. 

\section{MicroFog - Evaluation and Validation}\label{sec:evaluation}

In this section, we validate the main features and functions supported by MicroFog using multiple use cases. 
\subsection{Experimental Setup}
    \subsubsection{Infrastructure and MicroFog setup}
To evaluate the features supported by MicroFog, we create a prototype of a federated fog-cloud environment consisting of three Fog clusters (fog1, fog2 and fog3) and one Cloud cluster (cloud1). Each cluster belongs to a separate network and communicates with each other through load balancers. For the prototype, we use MetalLB \footnote{https://metallb.universe.tf/} as the load balancer that exposes each cluster to the outside. Each cluster is a separate Kubernetes cluster, and the communication among microservices running across different clusters is maintained by implementing an Istio service mesh across the clusters in multi-primary mode. Table \ref{table: infrastructure} summarises the details of each cluster. 

One of the main advantages of MicroFog is its compatibility with cloud-native technologies, which enables quick prototyping of federated fog-cloud architectures for placement algorithm development and evaluation to overcome the limitations due to the lack of publicly available Fog resources. To demonstrate this, we create the fog1, fog2 and fog3 clusters using virtualised resources available in the University of Melbourne's Queensberry Hall data centre, which is at the edge of the network and create cloud1 using AWS EC2 instances from ap-southeast-2 accessed through the internet. To replicate the behaviour of Fog clusters where Fog nodes are connected to each other through high bandwidth LAN links, we implement fog1, fog2 clusters as KinD Kubernetes (containerised k8s) clusters and fog3 as a k3d (containerised k3s) cluster belonging to separate sub-nets within the data centre. Their communication to the Cloud cluster occurs over the WAN network.

\begin{table}[t!]
	\caption{Federated fog-cloud infrastructure setup}
	\label{table: infrastructure}
	\centering
	\resizebox{0.6\linewidth}{!}
	{\begin{tabular}{| l | l| l|}
			\hline
			\multicolumn{1}{|c|}{\textbf{Cluster}} & 
                \multicolumn{2}{c|}{\textbf{Resources} }\\	
                \cline{2-3}
                \multicolumn{1}{|c|}{\textbf{Details}}& \textbf{CPU (VCPUs)} & \textbf{Memory (GB)} \\
                \hline 
			Cluster - fog 1 : & & \\
                node1 (control-node) & 3 & 6\\
                node2 (worker 1) & 4 &  9\\
                node3 (worker 2) & 5 & 16 \\
                node4 (worker 3) & 3 & 8\\
                \hline
                Cluster - fog 2 : & & \\
                node1 (control-node) & 3 & 6 \\
                node2 (worker 1) & 3 & 9\\
                node3 (worker 2) & 2 & 6\\
                node4 (worker 3) & 4 & 12\\
                node5 (worker 4) & 4 & 8 \\
                 \hline
                Cluster - fog 3 : & & \\
                node1 (server) & 3 & 6 \\
                node2 (agent 0) & 2 & 4\\
                node3 (agent 1) & 2 & 4\\
                 \hline
                Cluster - cloud 1 : & & \\
                node1 (control-node) & 8 & 14\\
                node2 (worker) & 8 & 14 \\
			\hline
	\end{tabular}}
    \vspace{-3.5mm}
\end{table}
    
\subsubsection{Workload Creation}
Due to the lack of diverse microservices-based IoT application benchmarks, we implement a tool to generate microservices-based mock applications \footnote{https://github.com/Cloudslab/MicroFog/tree/main/Workload\_Generator} that can capture different characteristics of MSA and generate heterogeneous applications for placement policy evaluation purposes. The tool provides a base microservice as a template that can be configured (using a Kubernetes ConfigMap) to create microservices that have multiple interaction patterns among them (i.e., chained, aggregate, or microservice candidate patterns) to create microservices-based applications having composite services that the users can access. Furthermore, the microservices created using the template can be configured to have different processing times and inter-microservice message sizes to fabricate the behaviour of heterogeneous applications. Using this tool, we create multiple microservices-based applications containing chained and aggregator interaction patterns to evaluate and verify different functionalities supported by the MicroFog framework.

\subsubsection{Placement Algorithm}
To highlight the main features supported by MicroFog, we adapt and implement different variations of the placement algorithm proposed in \cite{pallewatta2019microservices}. The algorithm in \cite{pallewatta2019microservices} aims to place the latency-critical IoT application services as close as possible to the user such that the resource requirements of the microservices are met. To this end, the placement policy starts placement from the traffic entry Fog clusters, moves towards adjacent Fog clusters and finally considers Cloud if the Fog resources are insufficient. We extend the policy in \cite{pallewatta2019microservices} to incorporate throughput awareness where the throughput of the composite services can be provided during PR submission, and the placement algorithm calculates the number of microservice instances and resources requirement to support the throughput. We use the calculation provided in \cite{pallewatta2022qos} for this. We create three variations of this approach to evaluate and validate multiple configurations and features of MicroFog as follows:
\begin{enumerate}[leftmargin=*]
    \item Version 1 (V1) - Vertically Scaled Distributed Placement: The placement algorithm retrieves already placed microservices from the PR and calculates the next microservice to place based on the DAG representation of the application. Afterwards, the algorithm tries to place the microservice within the cluster in a resource-aware manner. In this approach, since vertical scalability is considered, a single instance is placed for each microservice so that their resource allocation suffices the throughput requirement. If the cluster doesn't have enough resources to complete the application placement, the PR is updated and forwarded to the next cluster to place the rest of the microservices. 
    \item Version 2 (V2) - Horizontally Scaled Distributed Placement: This follows a similar approach to V1 but supports the horizontal scalability of the microservices. Thus, instead of a single instance, multiple instances of each microservice are placed to support the throughput requirement.
    \item Version 3 (V3) - Centralised Placement: In this version, the placement algorithm maintains a view of all available clusters. Once the request is received, the algorithm selects one of the entry clusters defined in the PR. Next, the algorithm traverses the DAG and places microservices starting from the selected Fog cluster, then consider adjacent clusters if no resources are available and finally considers Cloud for placement.

\end{enumerate}

\subsection{Use cases and results}

    \subsubsection{Analysing Flexibility and  Scalability of MicroFog Architecture}
    Flexibility and scalability of the MicroFog architecture is denoted by its ability to operate within distributed multi-fog  multi-cloud enviornments. We explore distributed deployment architecture of the MicroFog framework under different configurations to demonstrate this.
    
    \begin{itemize} [leftmargin=*]
    \item{ \textbf{Distributed Data management and access : }} 
    
    In this section, we analyse and validate the deployment architectures proposed in this paper for accessing MinIo Yaml File Store and Redis Meta Data Store. Our proposed deployment architectures aim to ensure lower latency and high availability of the data stores to ensure reliable placement and deployment of applications. To evaluate this, we consider three data access scenarios. Relative data retrieval latency is measured for each scenario as shown in Figure \ref{fig:datastores}(a) and Figure \ref{fig:datastores}(b) for MinIO YAML Store and Redis Meta Data Store, respectively. We submit placement requests to the CE placed in fog1 and observer behaviour under distributed placement mode. In Scenario 1, both data stores are deployed within all 3 clusters following the proposed architecture in Figure \ref{fig:yamlStore}. Scenario 2 considers the unavailability of fog1 data stores, whereas Scenario 3 considers the unavailability of data stores in both fog1 and fog2.

    Results demonstrate that the deployment architecture manages request routing to data stores as intended. The failover policy is configured to prioritise the closest data store in case of data store failures. Accordingly, if all data stores are available, the CE deployed within cluster fog1 accesses the data stored deployed within the same Fog cluster, thus resulting in the lowest data retrieval latency. If the data stores within the cluster are unavailable, the routing policy prioritises the closest adjacent Fog cluster over the Cloud cluster and only accesses the Cloud cluster in case the data stores in both Fog clusters are unavailable. This behaviour is depicted by the obtained latency values, which show a slight increase in latency due to failover triggered among Fog clusters (Scenario 2 - FO to Fog) and a relatively larger increase with failover from Fog to Cloud (Scenario 3 - FO to Cloud). Thus, the proposed deployment architecture is robust to ensure data access while aiming to improve performance. Furthermore, in the case of resource-constrained Fog clusters, it would be more feasible to host the data stores in adjacent resource-rich Fog clusters or Cloud clusters at the cost of data access performance. Our proposed architecture is flexible enough to support this behaviour and ensure data access across federated multi-fog multi-cloud environments.

\begin{figure}[!ht]
    \centering
    \includegraphics[width=\linewidth]{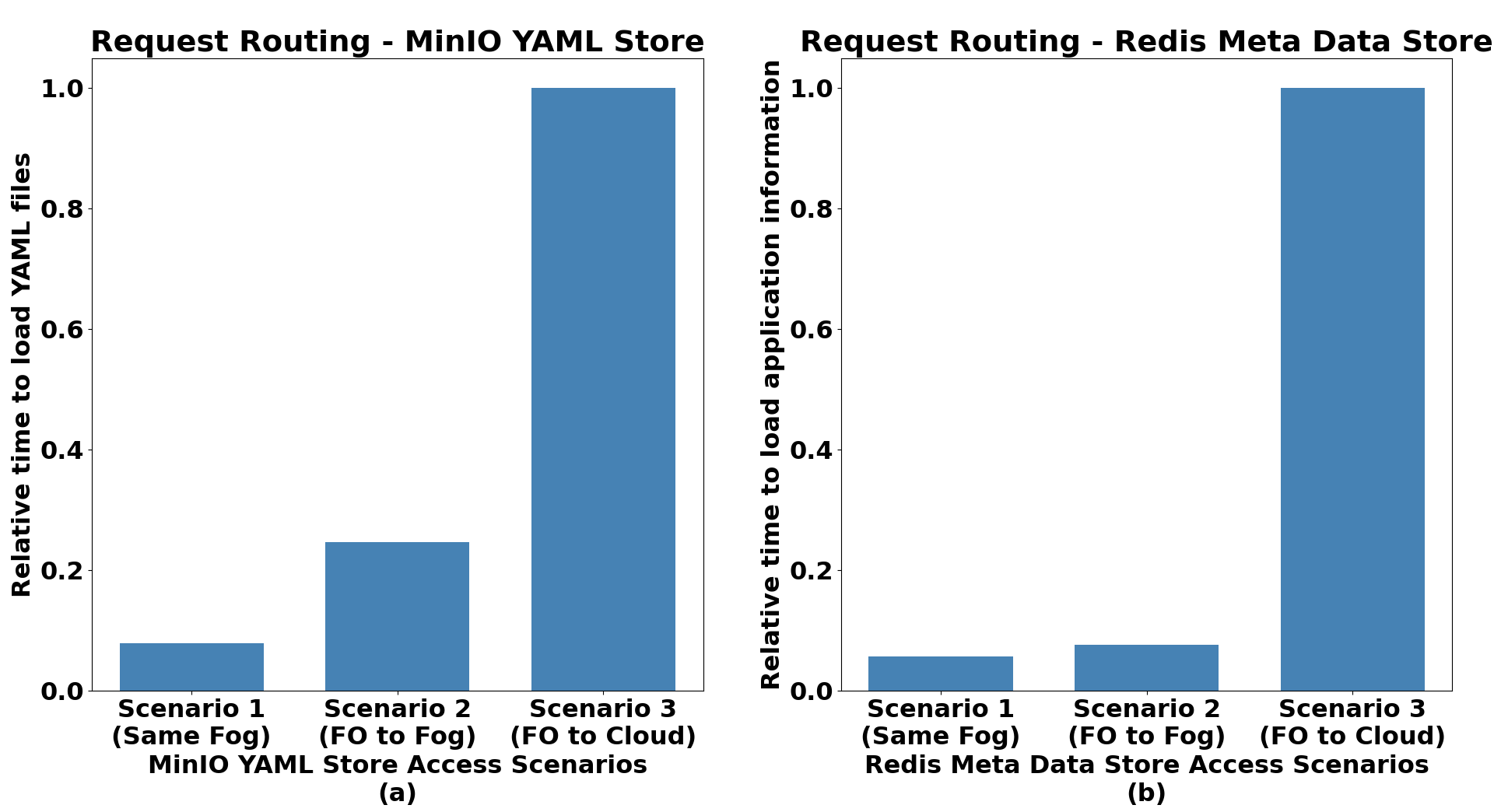}
    \vspace*{-5.0mm}
    \caption{Availability Analysis of Data Stores}
    \label{fig:datastores}
    \vspace*{-5.0mm}
\end{figure}


     \item{ \textbf{Analysis on Distributed Deployment of CE and its Operation Modes}}

    MicroFog-CE is designed for salable deployment across distributed Fog and Cloud clusters. To this end, CE supports distributed operation mode of the CE, where all CEs execute placement algorithms independently and the centralised mode, where the primary CE executes the placement algorithms and sends placement output to individual clusters. In both approaches connectivity among CEs are maintained using prososed deployment architecture (section \ref{sec:ceDeployment}) to achieve successful placement of applications.

    In the distributed mode, PRs can be forwarded to adjacent Fog or Cloud clusters, and MicroFog-CE supports the integration of different forwarding policies, thus providing the users of the framework with the flexibility to control distributed placement policies. We demonstrate this by implementing two forwarding policies, 1) FP1: if the current cluster does not have enough resources to complete PR placement, PR is forwarded to an adjacent Fog cluster, 2) FP2: if the current cluster does not have enough resources to complete PR placement, the PR is forwarded to a connected Cloud cluster. To route the PR to the selected cluster, the header of the PR forwarding request is updated with the destination cluster name. The deployment architecture proposed in Figure \ref{fig:controlengine} routes to the correct destination based on that. Figure \ref{fig:prForwarding} shows three scenarios where in Scenario 1, the entry Fog cluster for the PR contains enough resources to host the application, thus resulting in the lowest response time out of the three scenarios. Scenario 2 and Scenario 3 consider a situation where the entry Fog cluster does not have enough resources to host the entire application. Scenario 2 uses FP1, thus placing the application across two adjacent Fog clusters, which results in a higher response time than the prior scenario due to inter-fog communication delay. However, FP1 performs better than Scenario 3, which uses FP2, where the request is forwarded to the Cloud. This incurs the highest response time among the three scenarios. The above use case demonstrates the scalability of the CE deployment architecture to tackled multiple Fog and Cloud clusters and also the ability to configure distributed placement policies by integrating forwarding policies.  

\begin{figure}[!ht]
    \centering
    \includegraphics[width=0.75\linewidth]{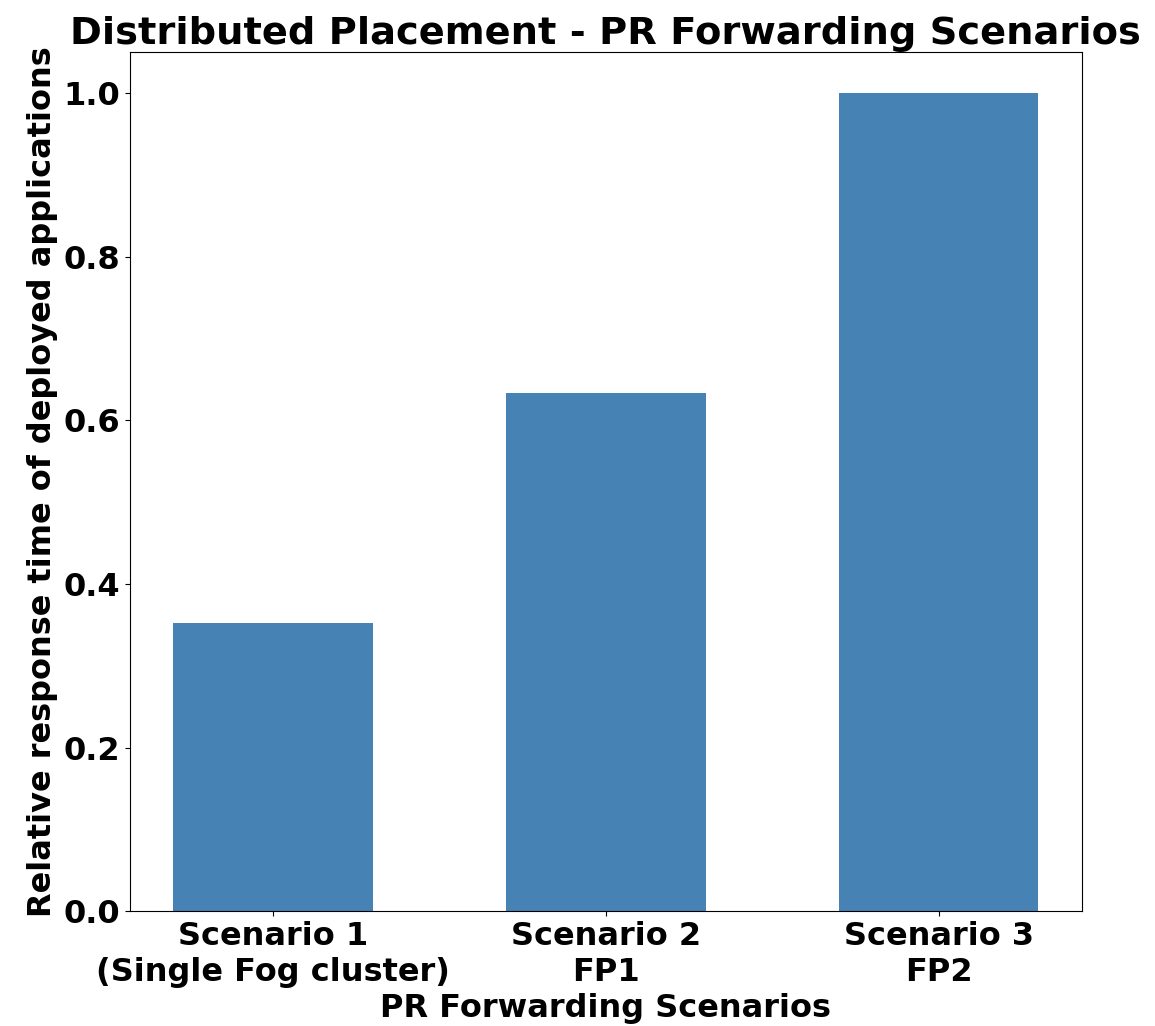}
    \caption{Distributed Placement Algorithm Execution}
    \label{fig:prForwarding}
    \vspace*{-5.0mm}
\end{figure}

    MicroFog-CE also supports centralised placement algorithm execution as well. In Figure \ref{fig:operatiomodes}, we consider three placement scenarios and analyse time to application placement under the CE's distributed and centralised operation mode. The three scenarios are as follows: Scenario 1 - 5 PRs are submitted to the system simultaneously such that three have \textit{fog1} as the entry cluster and the other two have \textit{fog2} as the entry cluster; scenario 2 - 10 PRs are submitted to the system simultaneously such that each receives 5PRs; scenario 3 - 15 PRs in total simultaneously submitted to fog1, fog2, fog3 such that each received 5 PRs. In the distributed operation mode PRs are submitted to the CE of their entry cluster, whereas in the centralised mode, all PRs are submitted to the primary CE deployed within the Cloud. Furthermore, the centralised mode uses V3, whereas distributed mode uses V2 as the placement policy. Figure \ref{fig:operatiomodes} depicts the total time for PR deployment, calculated from when the CE receives the PR to application deployment completion under event-driven placement mode. In all cases, distributed mode takes lesser time to complete application placement as more PRs are processed simultaneously. Thus the relative difference between completion time increases as the number of PRs increases. This demonstrates that depending on factors like the PR arrival rate, the design of the placement policy and the scale of the federated Fog environments, etc. MicroFog-CE can be configured to use centralised or distributed placement modes.

      \begin{figure}[!ht]
    \centering
    \includegraphics[width=0.75\linewidth]{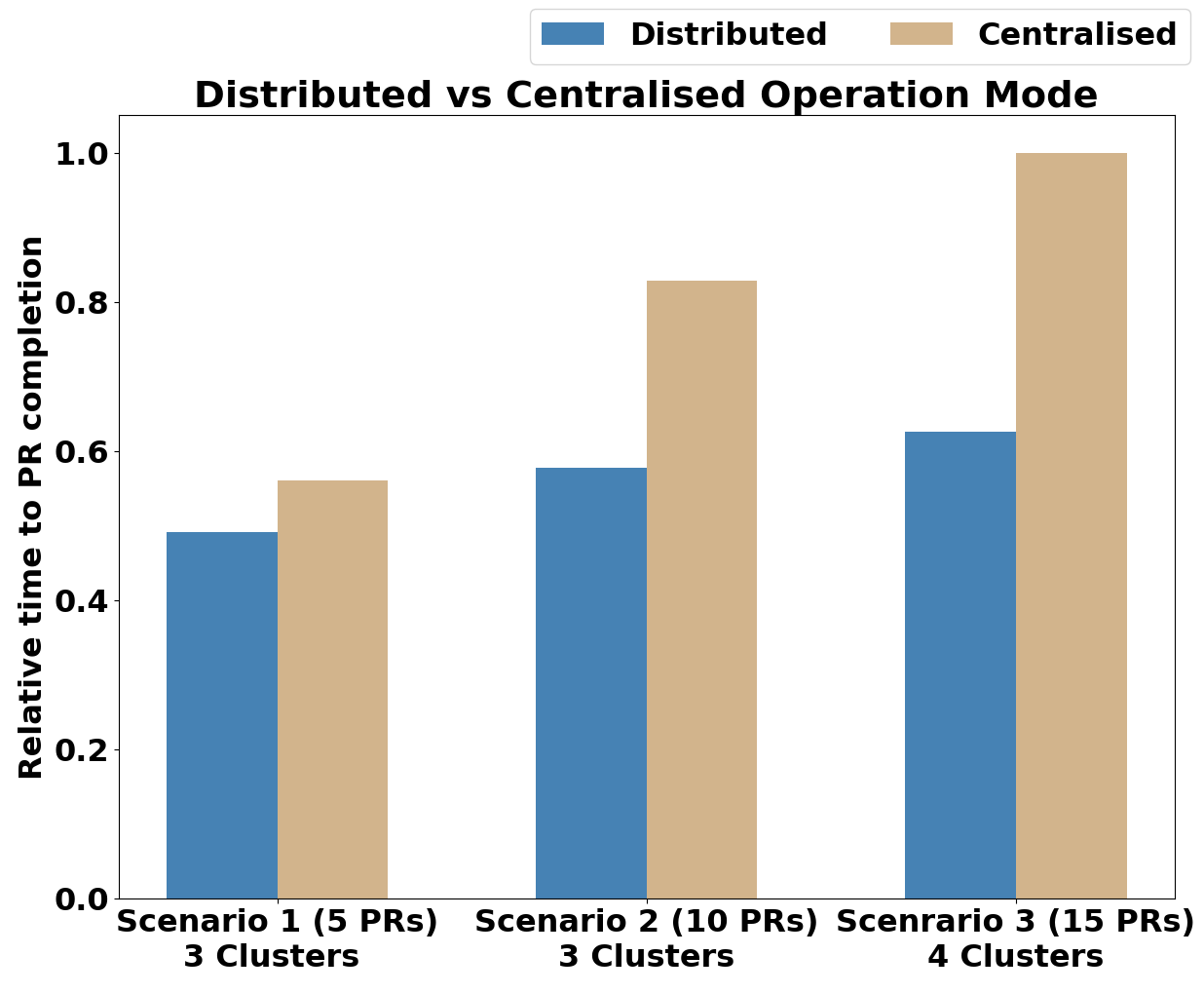}
    \caption{Analysis of CE Operation Modes}
    \label{fig:operatiomodes}
    \vspace*{-5.0mm}
\end{figure}

    \item{ \textbf{Analysis on Using Different Kubernetes Distributions}}

    Due to heterogeneous resource availability, Fog and Cloud clusters can run different Kubernetes distributions (i.e., k8s for resource-rich clusters and k3s for resource-constrained clusters). To analyse the ability of MicroFog to operate across different distributions. Results show that PR deployment time is lesser in fog3 (Scenario 2), which uses k3s due to its light architecture, whereas fog1 (Scenario 1) deployment time is higher. Furthermore, scenario 3 depicts a cross-cluster PR placement scenario, which takes longer than the k3s cluster but less time than the k8s deployment due to deployment across both. This demonstrates MicroFog-CEs' flexibility to operate across clusters with different Kubernetes distributions.  

\end{itemize}    
   \begin{figure}[!ht]
    \centering
    \includegraphics[width=0.75\linewidth]{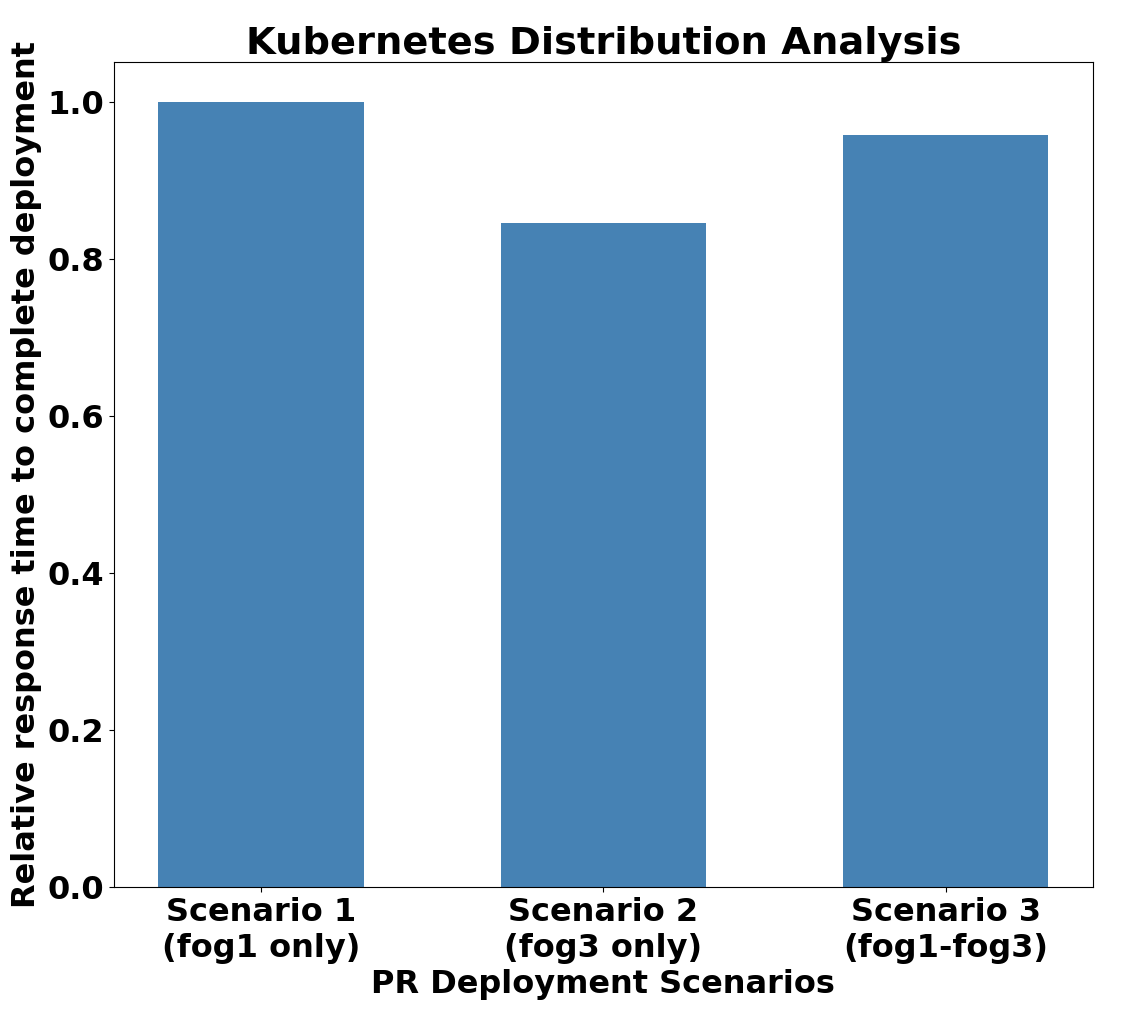}
    \caption{Analysis of Kubernetes Distributions}
    \label{fig:distributionComparison}
    \vspace*{-5.0mm}
   \end{figure}

    The above results demonstrate the ability of MicroFog to handle placement across multiple clusters (scalable architecture) and configurability (integration of different placement algorithms, forwarding policies, and operation modes) of the MicroFog-CE, which enables it to successfully execute placement policies and deploy applications across distributed Fog and Cloud clusters. 
    
    \subsubsection{Federated fog-cloud deployment and compositing (service discovery and load balancing) of microservices }

    One of the main advantages of MSA is the ability to independently scale microservices across distributed computing resources while ensuring their dynamic composition through service mesh technologies. As MicroFog-CE supports easy integration of multiple placement algorithms, we implement V1 and V2 to demonstrate the effect of scalable microservice placement and validate dynamic composition and load-balancing enabled by MicroFog. 

\begin{figure}[!ht]
    \includegraphics[width=\linewidth]{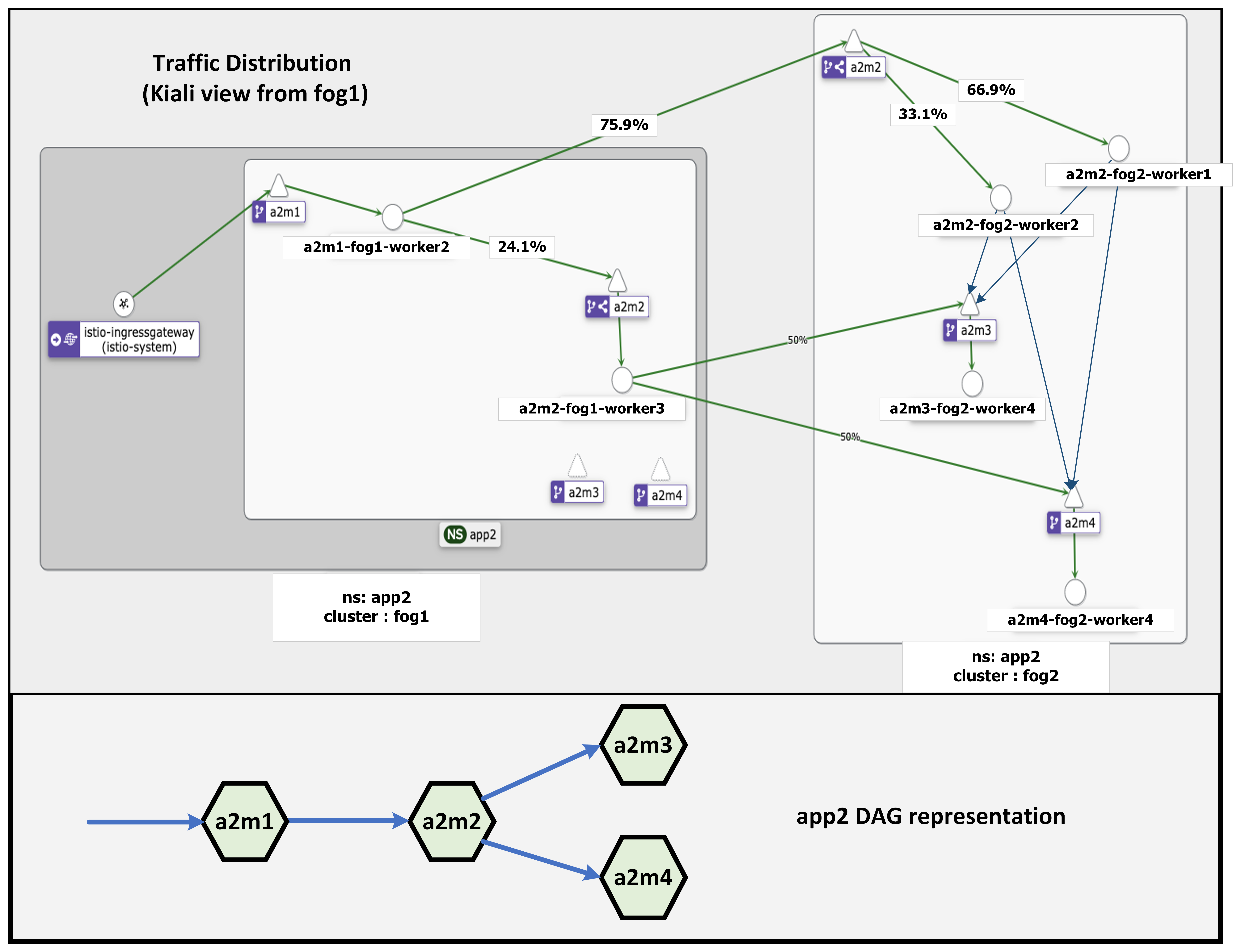}
    \vspace*{-4.0mm}
    \caption{Multi-cluster Service Discovery and Load Balancing Scenario - app2}
    \label{fig:loadbalancing}
    \vspace*{-2.0mm}
\end{figure}

\begin{figure}[!ht]
    \includegraphics[width=\linewidth]{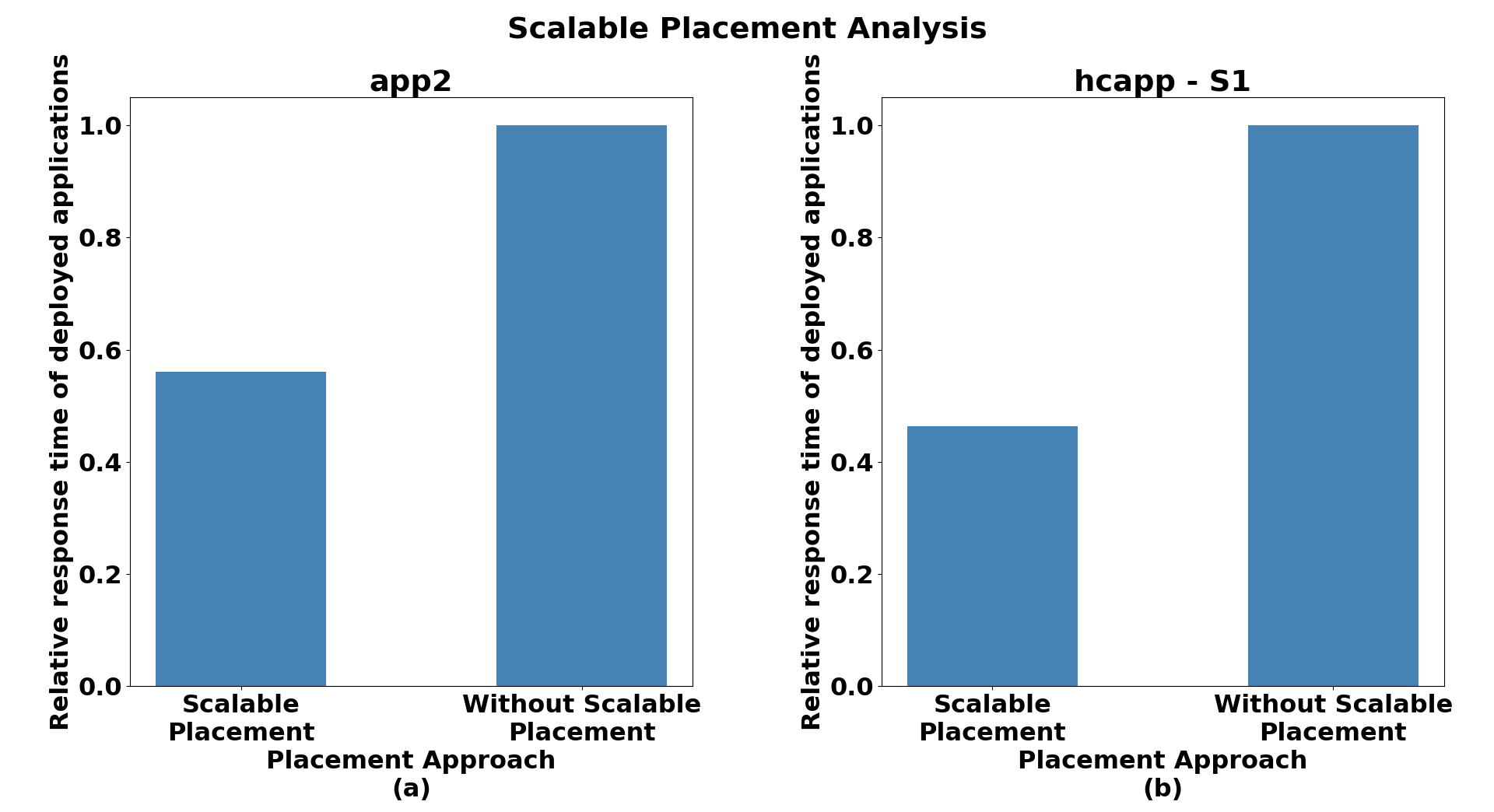}
    \caption{Scalable Microservice Placement}
    \vspace*{-5.0mm}
    \label{fig:scalableplacement}
    \vspace*{-2.0mm}
\end{figure}

    We consider the placement of two microservices-based applications generated using workload generator: smart healthcare application (application id: \textit{hcapp}) discussed as an example IoT application in Section \ref{sed:exampleDeployment} (see Figure \ref{fig:exapleAppDeployment}) consisting of two composite services, and a  DAG-based application (application id: \textit{app2}) which consists of a single composite service that can be accessed by the user (see Figure \ref{fig:loadbalancing}). The service consists of 4 microservices, where a2m1 and a2m2 form a chained invocation pattern and a2m2, a2m3, and a2m4 form an aggregator pattern such that a2m1 invokes a2m3 and a2m4, aggregates their results and return it back to a2m1 for further processing. The resultant placements generated by the two versions of the placement algorithm for \textit{app2} and \textit{hcapp} are recorded in Table \ref{table:loadbalancingT}. As V1 does not consider horizontal scalability, resource-constrained natures of the heterogeneous Fog nodes force the placement to move towards the Cloud, thus resulting in higher latency, as shown in Figure \ref{fig:scalableplacement}. In comparison to that, V2 utilises the ability to scale microservices horizontally. This results in better utilisation of limited Fog resources, thus resulting in lower latencies, as shown under scalable placement in Figure \ref{fig:scalableplacement}. Results demonstrate that, V2 improves latency by 44\% for \textit{app2} and 54\% for \textit{hcapp}.
    
\begin{figure}[!ht]
    \includegraphics[width=\linewidth]{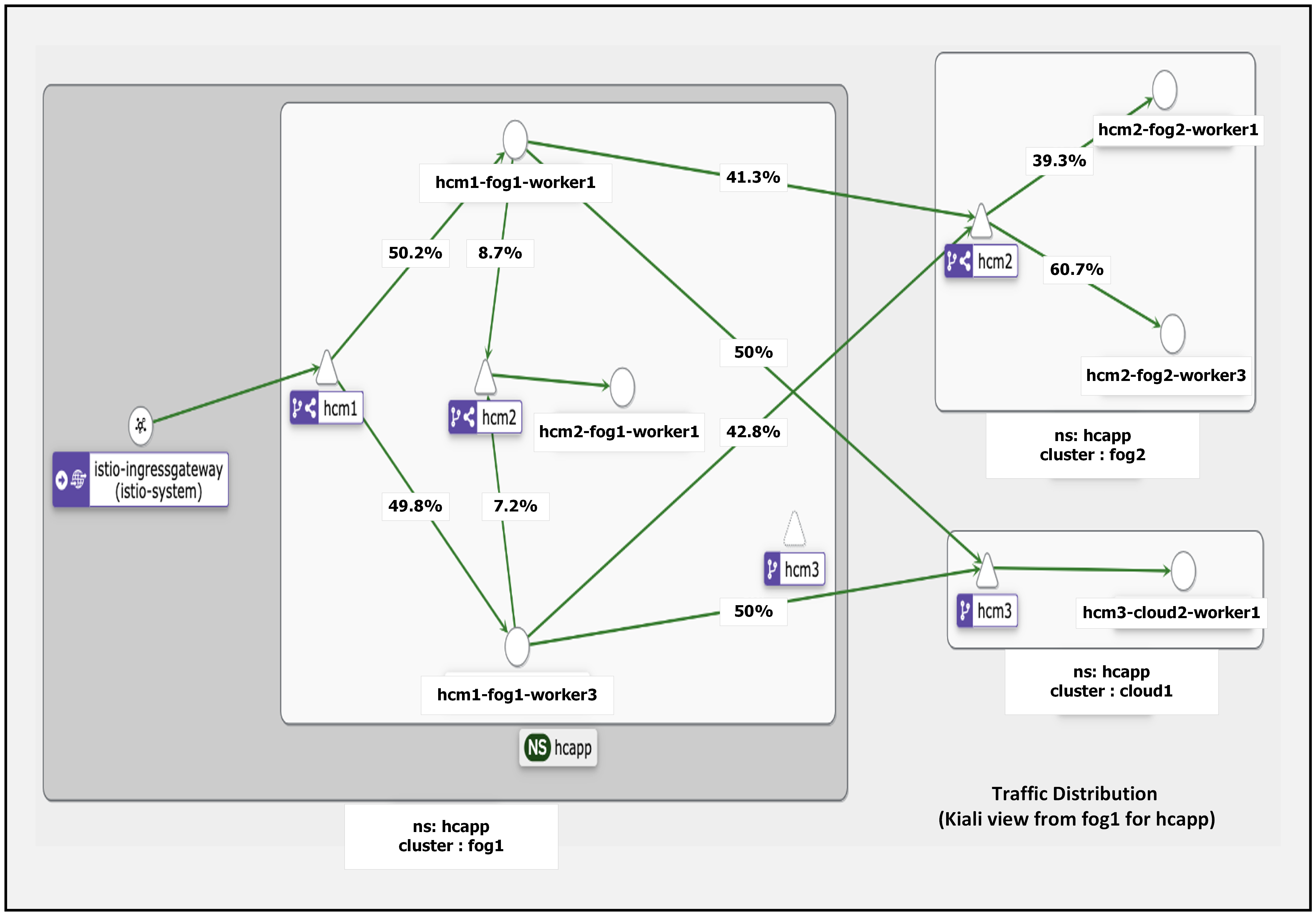}
    \vspace*{-4.0mm}
    \caption{Multi-cluster Service Discovery and Load Balancing Scenario - hcapp}
    \label{fig:loadbalancingHCAPP}
    \vspace*{-2.0mm}
\end{figure}

    However, dynamic service discovery and load balancing across clusters are required to ensure connectivity among microservices and maintain the expected level of performance. To this end, MicroFog-CE supports the integration of new load-balancing policies. In this experiment, we implement a Weighted Round Robin Load Balancing policy. Deployment rules of the MicroFog-CE deploy Istio VSs and DRs according to the output of the load balancing policy. For the above placement, we verify this based on the Kiali workload graph, which depicts the traffic distribution across different horizontally scaled instances of the same microservice. Table \ref{table:loadbalancingT} shows that for the horizontally scaled microservice a2m2 in \textit{app2}, the resource distribution is 1:2:1 among instances deployed within fog1-worker3, fog2-worker1 and fog2-worker2, respectively. Obtained graph (see Figure \ref{fig:loadbalancing} shows that traffic for a2m2 is divided with a 1:3 ratio among two clusters and 2:1 within the fog2 cluster, thus diving a2m2 traffic with an approximate ratio of 1:2:1 among its three instances. This matches with the expected traffic distribution of Weighted Round Robin load balancing, thus confirming the ability of the MicroFog to automate Istio resource deployment to ensure the custom load balancing capabilities across clusters. This is further demonstrated by Figure \ref{fig:loadbalancingHCAPP}, which reflects the traffic distribution of \textit{hcapp}. The traffic distributions of microservices hcm1 (1:1) and hcm2 (1:2:3) adheres to their resource distribution of hcm1 (1:1) and hcm2 (1:2:3).

    \begin{table}[t!]
	\caption{Generated Placement for example applications (app2 and hcapp)}
	\label{table:loadbalancingT}
	\centering
	\resizebox{\linewidth}{!}
	{\begin{tabular}{| c | c| c|c|c|}
                \hline
			\multicolumn{1}{|c|}{\textbf{Placement}} & 
               \multicolumn{2}{c|}{\textbf{app2}} &
                \multicolumn{2}{c|}{\textbf{hcapp} } \\
			\cline{2-5}
			\multicolumn{1}{|c|}{\textbf{Algorithm}} & 
               \multicolumn{1}{c|}{\textbf{Microservice}} &
                \multicolumn{1}{c|}{\textbf{Deployed Nodes} } &
                \multicolumn{1}{c|}{\textbf{Microservice}} &
                \multicolumn{1}{c|}{\textbf{Deployed Nodes} }\\
                \hline
			& a2m1 &  fog1-worker2  & hcm1 & fog2-worker3\\
                \cline{2-5}
                Version 1 & a2m2 & fog2-worker4 & hcm2 & cloud1-worker1  \\
                \cline{2-5}
                (V1)& a2m3 & cloud1-worker1 & hcm3 & cloud1-control-node \\
                \cline{2-5}
                & a2m4 & cloud1-worker1 & &  \\
                \hline
                
                & a2m1 & fog1-worker2 & hcm1 & fog1-worker1, fog1-worker3 \\
                \cline{2-3}
                Version 2 & a2m2 & fog1-worker3, fog2-worker1, fog2-worker2  &  & Allocated Resource Ratio - 1:1 \\
                \cline{4-5}
                (V2) &  & Allocated Resource Ratio - 1:2:1 & hcm2 & fog1-worker1, fog2-worker1, fog2-worker3\\
                \cline{2-3}
                & a2m3 & fog2-worker3  &  &  Allocated Resource Ratio - 1:2:3\\
                \cline{2-5}
                & a2m4 &  fog2-worker4 & hcm1 & cloud1-control-node\\
			\hline
	\end{tabular}}
    \vspace{-3.5mm}
\end{table}

  Results obtained from the above use cases capture different features supported by MicroFog
   and verify that MicroFog is a scalable and easy-to-configure framework that can deploy microservices across federated Fog computing environments and ensure dynamic microservice composition across clusters. Hence, the MicroFog framework can be successfully used and extended for integrating and evaluating the performance of novel placement algorithms designed for the placement of microservices-based IoT applications.



\section{Conclusions and Future Work}\label{sec:conclusion}

In this work, we proposed a framework for the scalable placement of microservices-based IoT Applications in federated Fog environments. The proposed framework is scalable, extensible and configurable to execute placement algorithms and deploy applications across Kubernetes and Istio-enabled multi-fog multi-cloud environments. Moreover, the framework provides the ability to integrate novel placement policies, load balancing policies and PR forwarding policies. Thus, placement algorithm developers and IoT application developers can use the framework to deploy their applications within federated Fog environments and monitor their performance. Furthermore, the framework provides rapid prototyping support, and the applications developed following MSA do not require any application-level changes to be deployed using the framework. Thus, the framework abstracts the underlying deployment-related functionalities from the users, giving them a chance to focus more on placement policy development and IoT application development. 

Due to the use of open-source technologies, modular design and architecture, developers can easily extend the framework to add novel functionalities. As future work, the MicroFog framework can be further improved with lightweight security mechanisms for data transmission across clusters, a scalable architecture to store and use observability-related data to improve placement algorithms, ability to integrate novel fault-tolerance policies for applications.

	
	%

	

	\section*{Software Availability}
	
The source code and documentation of the MicroFog framework is accessible from: \newline \href{https://github.com/Cloudslab/MicroFog}{https://github.com/Cloudslab/MicroFog}

\section*{Acknowledgements}
         We thank Melbourne Research Cloud (MRC) for providing the infrastructure used for implementing the MicroFog prototype.
	\bibliographystyle{IEEEtran}
	
	\bibliography{reference}
	
\end{document}